\documentclass[%
 reprint, amsmath, amssymb, aps, superscriptaddress, prx]{revtex4-2}
\usepackage{graphicx}% Include figure files
\usepackage{bm}% bold math
\usepackage{siunitx}
\usepackage{soul} 
\usepackage{braket}
\usepackage{setspace}
\usepackage{bm, dsfont}% bold math
\usepackage[dvipsnames]{xcolor} 
\usepackage[colorinlistoftodos]{todonotes}
\usepackage[colorlinks=true, linkcolor=NavyBlue, citecolor=NavyBlue, urlcolor=NavyBlue, breaklinks=true]{hyperref}
\usepackage{color}
\usepackage[utf8]{inputenc}
\usepackage{ulem}
\usepackage{soul}
\usepackage{epstopdf}

\newcommand{\rmd}{{\rm d}}
\newcommand{\rmi}{{\rm i}}
\newcommand{\e}{{\rm e}}
\newcommand{\gammarad}{\Gamma_{1}}

\begin{document}

\title{Tunable directional photon scattering from a pair of superconducting qubits}

\author{Elena S. Redchenko}
\email{elena.redchenko@ist.ac.at}
\affiliation{Institute of Science and Technology Austria, 3400 Klosterneuburg, Austria}
\author{Alexander V. Poshakinskiy}
\affiliation{Ioffe Institute, St.~Petersburg 194021, Russia}
\author{Riya Sett}
\affiliation{Institute of Science and Technology Austria, 3400 Klosterneuburg, Austria}
\author{Martin Zemlicka}
\affiliation{Institute of Science and Technology Austria, 3400 Klosterneuburg, Austria}
\author{Alexander N. Poddubny}
\affiliation{Ioffe Institute, St.~Petersburg 194021, Russia}
\author{Johannes M.~Fink}
\email{jfink@ist.ac.at}
\affiliation{Institute of Science and Technology Austria, 3400 Klosterneuburg, Austria}
\date{\today}

\begin{abstract}
The ability to control the direction of scattered light 
%in integrated devices 
is crucial to provide the flexibility and scalability for a wide range of on-chip applications, such as integrated photonics, quantum information processing and nonlinear optics. 
In the optical and microwave frequency ranges tunable directionality can be achieved  by  applying external magnetic fields that modify  optical selection rules~\cite{Sllner2015,Spitzer2018}, by using nonlinear effects~\cite{hamann2018nonreciprocity}, or interactions with vibrations~\cite{poulton2012design, kittlaus2017chip, safavi2019controlling}.  However, these approaches are less suitable to control propagation of microwave photons inside integrated superconducting quantum devices~\cite{Reuer2022}, that is highly desirable.
%in the microwave frequency range
%Nonreciprocal devices can be realized using magnetic field \cite{zvezdin1997modern}, nonlinear effects \cite{hamann2018nonreciprocity}, stimulated Brillouin scattering \cite{poulton2012design, kittlaus2017chip}, or parametrically driven mechanical systems \cite{barzanjeh2017mechanical, safavi2019controlling}. 
Here, we demonstrate on-demand tunable directional scattering based on two periodically modulated transmon qubits coupled to a transmission line at a fixed distance \cite{astafiev2010resonance,vanLoo1494,Kannan2022} in close analogy to two oscillating mirrors \cite{poshakinskiy2019optomechanical,mirhosseini2019cavity}. By continuously changing the 
%symmetry of the modulation, governed by 
relative phase 
%$\alpha$ 
between the local modulation tones, 
we realize unidirectional forward or backward photon scattering. Such an in-situ switchable mirror represents a versatile tool for intra- and inter-chip microwave photonic processors. In the future, a lattice of qubits can be used to realize topological circuits that exhibit stronger nonreciprocity or chirality \cite{Tomoki2019}.
\end{abstract}

\maketitle

One of the simplest ways to realize directional light scattering relies on the Kerker effect~\cite{kerker1983electromagnetic,Liu2018}. It is based on the interference between different multipole components of scattered light, for example electric and magnetic dipoles, and has been demonstrated for Si nanoparticles~\cite{Staude2013,fu2013directional,person2013demonstration}. However, the nanoparticle scattering pattern is fixed after fabrication and dictated by its shape. Tunable light routing is typically enabled by an external magnetic field that leads to the Zeeman splitting of optical transitions for clockwise- and counter-clockwise- propagating photons~\cite{Sllner2015} or a modification of optical selection rules~\cite{Spitzer2018}. In the optical domain, the routing can be reversed as well without changing the magnetic field by flipping the spin of the atom \cite{scheucher2016quantum}. Such structures are now actively studied in the domain of chiral quantum optics~\cite{lodahl2017chiral,Prasad2020}. Tunable directional scattering can also be achieved by using moving boundary conditions~\cite{poshakinskiy2019optomechanical,zhang2018thermal}.   
For example, the trembling of a small particle with only an electric dipole resonance can induce a magnetic dipole resonance~\cite{poshakinskiy2019optomechanical}, which in turn provides directional scattering in analogy to the Kerker effect. Several compact nonmagnetic realizations of nonreciprocal devices using Raman and Brilliouin scattering \cite{muller2010measurement, poulton2012design, kittlaus2018non,safavi2019controlling} have been presented for optical frequencies. 

Tunable directional interactions are also highly desired for superconducting quantum circuits in the microwave spectral range. For instance, isolators and circulators are commonly used for superconducting quantum computing to protect fragile qubits states. Cascaded photon processing in the chiral setup is also beneficial for the creation of complex entangled quantum states of qubits~\cite{Guimond2020}. However, it is difficult to directly extend existing approaches for visible light to microwave photons. 
For example the classical Kerker approach is not applicable to a typical transmon qubit that behaves just as an electric dipole~\cite{astafiev2010resonance}, without magnetic dipole resonances.
Devices, based on an external magnetic field,~\cite{mahoney2017chip} are often bulky and always require additional shielding to protect superconducting qubits. While there exist demonstrations of directionality in parametrically driven, compact mechanical systems \cite{peterson2017demonstration,bernier2017,barzanjeh2017mechanical}, integration with superconducting circuitry is challenging due to limited bandwidth and tunability. Thus, there is a need for  flexible to use on-chip microwave photon routers which do not require strong magnetic fields or moving mechanical parts.

%%%%%%%%%%%%%%%%%%%%%%%%%%%%%%%%%%%%%%%%%%%%%%%%%%%%%%%%%%%
\begin{figure}[t]
\centering
\includegraphics[width=\columnwidth]{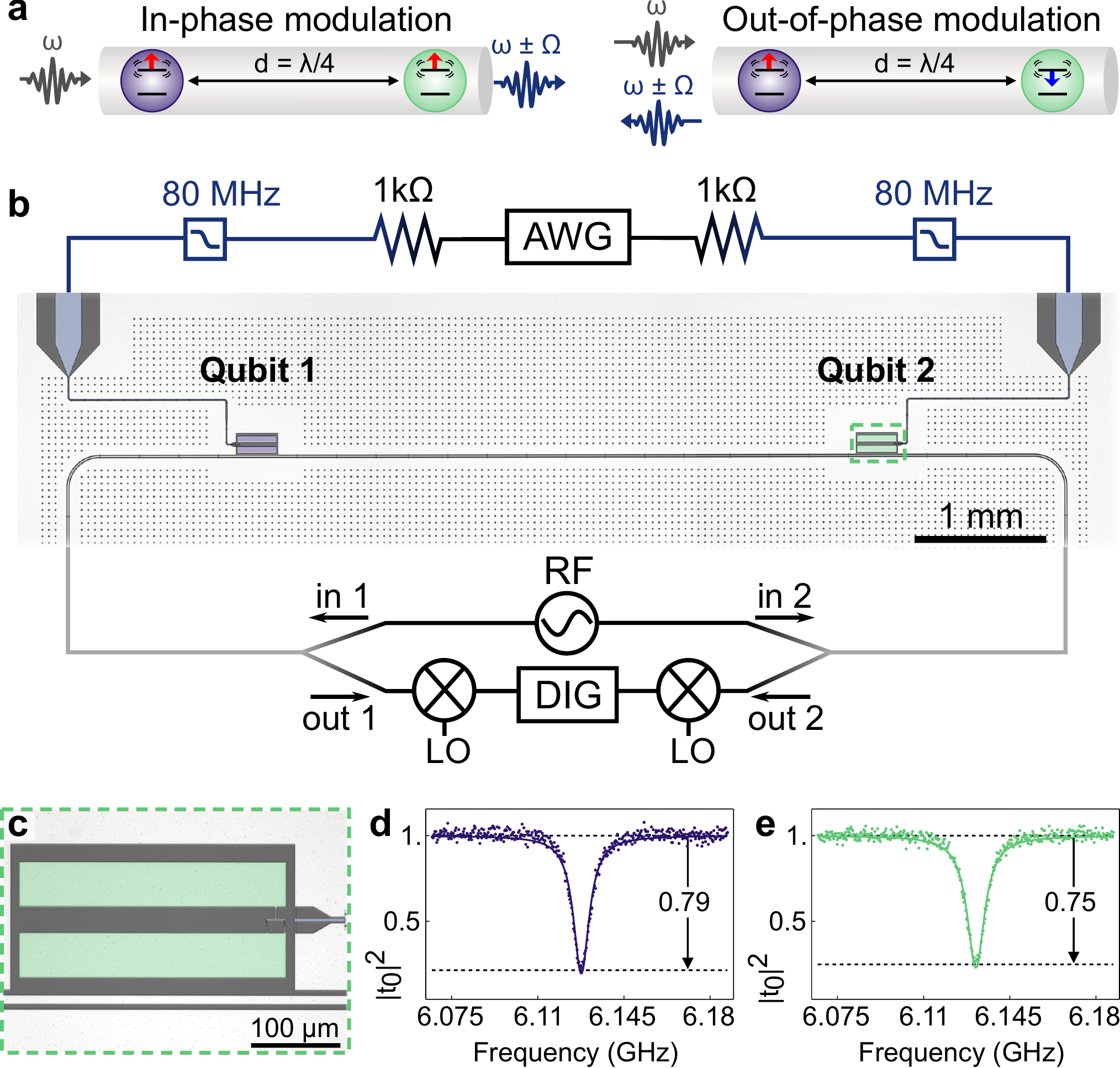}
\caption{\textbf{Experimental realization. a,} Schematic showing the scattering direction of the $\omega\pm\Omega$ component for in-phase (up-up) and out-of-phase (up-down) modulation of the qubits' transition frequencies $\omega$. \textbf{b,} Optical microscope image and simplified experimental setup. Two transmon qubits are capacitively coupled to a 50\,$\Omega$ transmission line, and each qubit has a local flux bias line connected to an arbitrary waveform generator channel (AWG), which is used to generate a sinusoidal wave with an amplitude $A_V$ that is filtered with a 80 MHz low pass filter and applied to ground via a 1\,k$\Omega$ resistor. We use an RF source, analog downconversion 
%(LO) 
and digitization (DIG) to back out the scattering parameters of the device cooled to 10\,mK.
%The device is mounted on the mixing chamber plate of a cryogen-free dilution refrigerator at a temperature $\sim 10$ mK. 
\textbf{c,} Enlarged view of Qubit 2 and local flux bias line inductively coupled to the qubit SQUID. \textbf{d (e),} Individually measured and normalized transmission spectra $|t_0|^2$ of elastically scattered radiation from Qubit 1(2) with fit to theory (solid line).} \label{fig1}
\end{figure}
%%%%%%%%%%%%%%%%%%%%%%%%%%%%%%%%%%%%%%%%%%%%%%%%%%%%%%%%%%%%

Here, our goal is to demonstrate an easy-to-fabricate circuit providing frequency and directionality tunable photon scattering with the minimum number of components required. 
%we actually also tune the scattering direction in-situ and the banwidth can be changed, as well as the suppression strength.
Our approach is based on the sinusoidal time-modulation of the qubit frequency \cite{chapman2017widely, rosenthal2017breaking, Shadrivov2018}, which is a standard technique in circuit and waveguide QED. The modulated qubit strongly coupled to a waveguide can be mapped onto the problem of light scattering from the trembling mirror \cite{jackel1992fluctuations,poshakinskiy2019optomechanical}.
By altering the relative phase $\alpha$ between the modulation tones of two qubits, we change the effective phase shift between the scattered sidebands resulting in different interference patterns for forward and backward scattering as schematically shown in Fig.~\ref{fig1}a.   
Here we do not focus on the elastic scattering nonreciprocity~\cite{chapman2017widely, rosenthal2017breaking} or directional emission from the initial qubit state~\cite{oliver2020b,Gheeraert2020} but on the switching between forward and backward inelastic coherent scattering. Thus, although elastically (Rayleigh) scattered radiation remains almost unaffected, we gain the flexibility to choose the frequency of the scattered photons.

\vspace{0.25cm}
%\noindent
%\textbf{Results}\\
\noindent
\section*{Experimental implementation}
We fabricate the sample with two transmon qubits coupled to a 1D coplanar transmission line separated by $d=5$\,mm  as shown in Fig.~\ref{fig1}b. The maximum frequency of the $|0\rangle \rightarrow |1\rangle$ transition is $9.129$ ($9.577$) GHz for Qubit 1(2). We tune both qubits to $\omega_0/(2\pi) = 6.129$ GHz corresponding to an effective distance of $d = \lambda/4$, with $ \lambda$ the wavelength of photons at  $\omega_0$, using bias coils mounted on top of the sample box. Currents for the periodic frequency modulation are applied via on-chip bias lines inductively coupled to the SQUID loops as shown in Fig. \ref{fig1}c. Both ports of the transmission line are connected to separate microwave in- and output lines to measure reflection and transmission spectra simultaneously.

Firstly, we characterize the qubits individually at $\omega_0$ where $d=\lambda/4$ using a weak resonant probe tone and measuring the coherently and elastically scattered radiation, i.e. at the same frequency. We determine the normalized transmission spectrum of each qubit shown in Fig.~\ref{fig1}d(e) and find the radiative decay rates to be $\gammarad / (2\pi) \approx 4.4$ MHz 
%at $6.129$ GHz 
with the dephasing rates of $\Gamma_{2} / (2\pi) \approx 3.9$ $(4.3)$ MHz for Qubit 1 (2) which is dominated by flux noise due to the relatively high flux dispersion at this bias point. 
%at $\omega_0$.
%maximum transition frequency. 
%of \hl{xyz (xyz) GHz}.
%Relatively high dephasing can be reduced by fabricating qubits with lower maximum transition frequency. However, it would then complicate the frequency modulation.

\begin{figure*}[t]
\centering
\includegraphics[width=1.5\columnwidth]{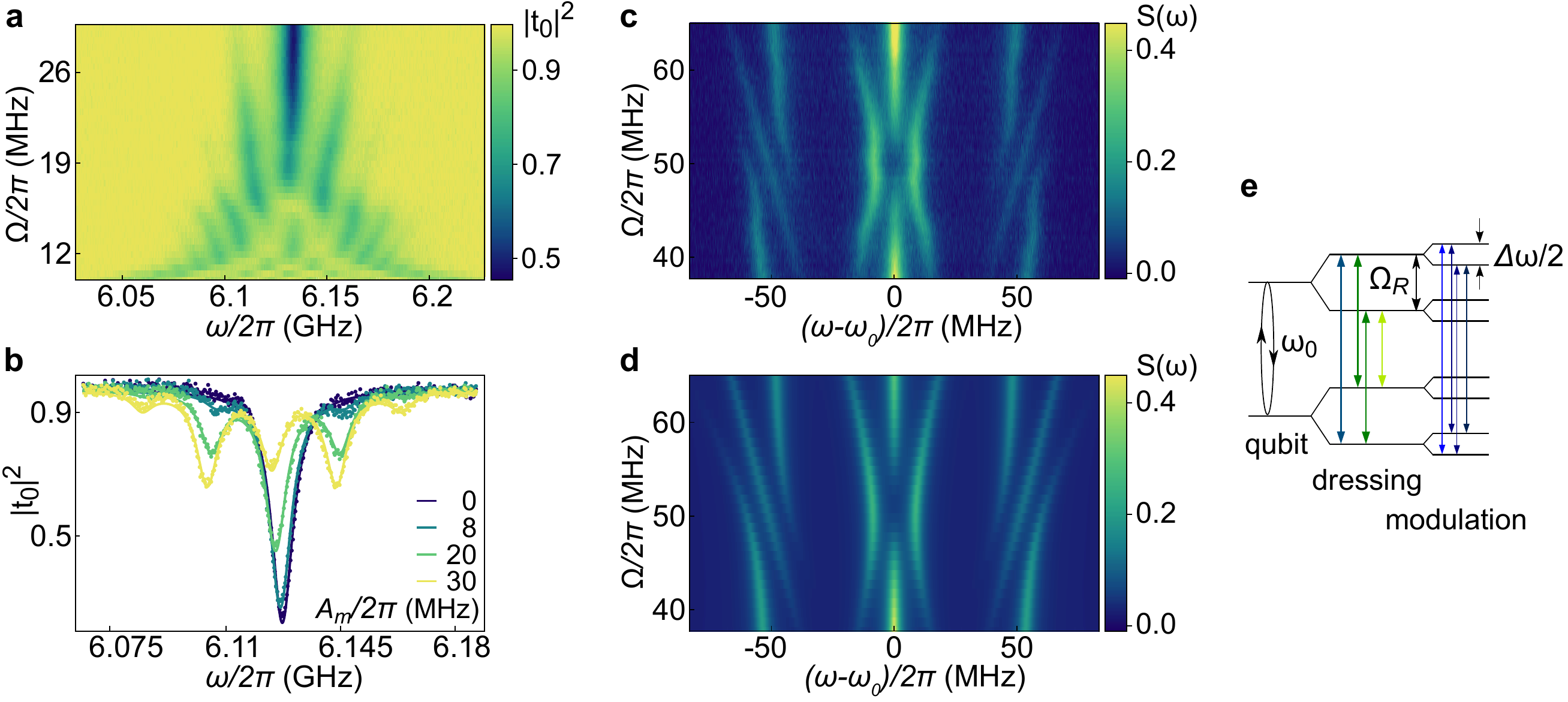}
\caption{\textbf{Singe qubit properties. a}, Normalized transmission spectrum $|t_0|^2$ of qubit 1 measured as a function of the modulation frequency $\Omega$ and the probe frequency $\omega$ at the fixed modulation amplitude $A_V = 50$\,mV$_\text{pp}$. \textbf{b}, Measured $|t_0|^2$ of the modulated qubit with $\Omega/(2\pi)=20$\,MHz for different $A_m$ and fits to Eq.~\ref{coh_oneQ} (solid lines). \textbf{c}, Measured resonance fluorescence emission spectrum of qubit 1 as a function of the modulation frequency $\Omega$ and detuning of the detected inelastically scattered radiation from the drive applied at $\omega_0$ for a Rabi frequency $\Omega_\text{R}/(2\pi)=52$\,MHz and modulation amplitude $A_m=0.2\,\Omega_\text{R}$. \textbf{d}, Theoretically predicted Mollow spectrum in the presence of frequency modulation for the same parameters. \textbf{e}, Level splitting schematics of the dressed and modulated qubit, which explains the origin of the observed nested Mollow triplets at $\Omega=\Omega_R$.
} \label{fig2}
\end{figure*}

%%%%%
%%%%%
%To further characterize the system, we study qubit spectra in the presence of modulation. 
%Sinusoidal modulation tones are generated by an arbitrary waveform generator (AWG). 
%Two outputs of the AWG are connected in series with 1 k$\Omega$ resistors at room temperature creating oscillating current with an amplitude $A_{V}$ which are then applied to the DC-lines. We use $80$~MHz low-pass LC filters at the mixing chamber (MC) to reduce the flux noise. Thus, the side-bands we can generate are limited to the maximum of $\pm 80$ ~MHz. 
An applied sinusoidal bias current makes the qubit resonance frequency tremble in time and the coherent %\hl{normalized? $|t_0|$} 
transmission amplitude 
%in the presence of one modulated qubit 
is then given by
\begin{eqnarray}\label{coh_oneQ}
&& t_0 = 1 + \sum_{n=-\infty}^{\infty}{\frac{i\gammarad/2}{\omega_0+n\Omega-\omega-i\Gamma_{2}}J_{n}^2\left( \frac{A_m}{\Omega}\right)},
\end{eqnarray}
where $J_{n}\left( \frac{A_m}{\Omega}\right)$ are Bessel functions of the first kind, $A_m$ is the modulation amplitude in frequency units, and $\Omega$ is the modulation frequency.
We measure the normalized transmission spectrum $|t_0|^2$ as a function of modulation frequency $\Omega$ as shown in Fig.~\ref{fig2}a. For the fixed signal amplitude at the AWG output $A_V = 50$\,mV$_\text{pp}$, the system undergoes a transition from the strong ($A_m/2>\Omega$) to the weak ($A_m/2<\Omega$) modulation regime. We fit similarly measured data to Eq.~\ref{coh_oneQ} for both qubits individually, as shown in Fig.~\ref{fig2}b for different applied $A_V$ and $\Omega/ (2\pi)=20$\,MHz and find that the fitted $A_m$ scales approximately linearly with $A_{V}$, which is shown in the Supplementary Information.
%%%%
%%%%

%%%%%%%%%%%%%%%%%%%%%%%%%%%%%%%%%%%%%%%%%%%%%%%%%%%%%%%%%%%
%%%%%%%%%%%%%%%%%%%%%%%%%%%%%%%%%%%%%%%%%%%%%%%%%%%%%%%%%%%%

\vspace{0.25cm}
%\noindent
%\textbf{Results}\\
\noindent
\section*{Modulated Mollow resonance fluorescence}
%Additionally, we study the resonance fluorescence spectrum of a single qubit in the presence of frequency modulation. It is well known that the spectrum of the inelastically scattered radiation from a superconducting qubit measured with the resonant drive at power $\Omega_{\rm R}> \gammarad$ demonstrates a Mollow triplet~\cite{astafiev2010resonance}. 
One of the hallmark characteristics of quantum two-level systems is the observation of the incoherent resonance fluorescence spectrum taking the form of a Mollow triplet for an applied resonant drive of sufficient power $\Omega_{\rm R}> \gammarad$~\cite{astafiev2010resonance}. Here we observe this effect for a frequency modulated qubit with $A_m = 0.2\,\Omega_{\text R}$ and Rabi frequency $\Omega_{\text R}/(2\pi) = 52$ MHz. The measured power spectral density (PSD) as a function of the modulation frequency $\Omega$ 
%and detuning of the detected radiation 
is shown in Fig.~\ref{fig2}c and the corresponding theory in Fig.~\ref{fig2}d. Dressing with the drive leads to the well known emission spectrum with three maxima at $\omega_0$ and $\omega_0\pm \Omega_R$. However, the additional frequency modulation leads to
%to an additional splitting of the regular Mollow triplet maxima and 
the formation of avoided crossings at $\Omega=\Omega_R$,
%at $\omega_0$ and $\omega_0\pm \Omega_R$. These avoided crossings 
which can be qualitatively interpreted as a formation of nested Mollow triplets following the level scheme shown in Fig.~\ref{fig2}e. Specifically, each of the levels of the original Mollow triplet is split into two levels due to the modulation. Next, the photon transitions between the split levels lead to the formation of additional Mollow triplets. For example, the transition from the original triplet having the largest energy, and shown by the thick vertical blue arrow, is transformed by the modulation into three distinct transition energies shown by the thin blue lines. The observed splitting between the outermost transitions of the inner Mollow triplets for $\Omega=\Omega_R$ is equal to $\Delta\omega/(2\pi)\approx 20$\,MHz, in excellent agreement with the numerical calculation. 

Similar formations of nested Mollow triplets in the electron spin-noise spectrum have been predicted for the conditions of electron paramagnetic resonance when the electron is subject to a the time-modulated magnetic field \cite{poshakinskiy2020spin}, but have so far not been observed directly to the best of our knowledge. 

%The  quantitative agreement between our measured results for the qubits with time-modulated resonance frequency and the theoretical calculations demonstrates the high setup quality.  This allows us to proceed with confidence to consider directional light scattering from the two-qubit structure.

\vspace{0.25cm}
\noindent
\section*{Directional scattering}
\begin{figure}[t]
\centering
\includegraphics[width=\columnwidth]{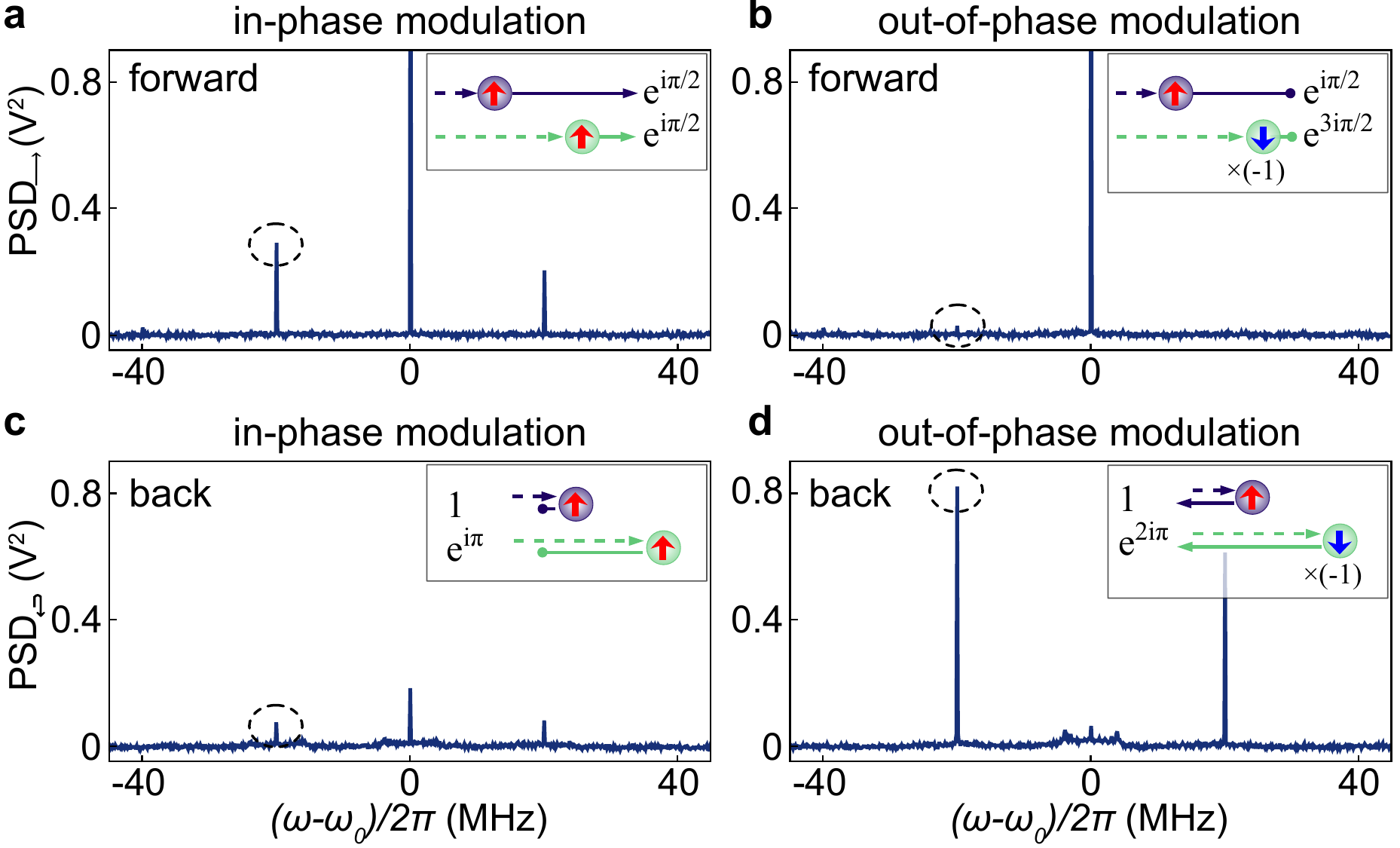}
\caption{\textbf{Resonance fluorescence spectra.} Power spectral density (PSD) measured in transmission (\textbf{a, b}) and reflection (\textbf{c, d}) at the digitizer for in-phase $\alpha = 0$ (\textbf{a, c}) and out-of-phase $\alpha = \pi$ (\textbf{b, d}) modulation. The Stokes components are highlighted with dashed circles. Scattering schematics are shown as insets where blue (green) arrows represent the light scattered from qubit 1 (2) at $\omega_0 \pm \Omega$ leading to constructive interference in \textbf{a} and \textbf{d} or destructive interference in \textbf{b} and \textbf{c}. Full Rayleigh peak heights are 1.9 and 1.6\,$V^2$ for the chosen settings in \textbf{a} and \textbf{b}.} \label{fig3}
\end{figure}

Now we consider the system of two qubits both tuned to $\omega_0$ and located at a distance $\lambda/4$. For any odd multiple of $\lambda/4$ a single resonant microwave tone drives the two qubits with opposite phase, which leads to a coherent exchange interaction mediated by virtual photons \cite{vanLoo1494} forming a coupled two-qubit molecule \cite{Kannan2022}. In the absence of modulation the backscattering is suppressed by destructive interference~\cite{Guimond2020}, while the interference for forward scattering is constructive.
The addition of frequency modulation of both qubits with $\Omega/(2\pi)=20$\,MHz and $A_m/(2\pi) = 20$\,MHz results in nontrivial interference conditions for the Stokes and anti-Stokes side-bands, as shown in the insets of Fig.~\ref{fig3}. 
%specifying the scattering direction, by the arrows in the insets of Fig.~\ref{fig3}. 
The blue and green arrows correspond to the incident light (dashed) and the inelastically scattered light (solid) at $\pm 20$ MHz from the first and second qubit, respectively. 
%The length of the arrows shows the gained phase. 
%The total length thus describes the overall scattering process.
If the two modulation tones are in phase ($\alpha=0$), illustrated in the insets of Fig.~\ref{fig3}a and c by red arrows inside the qubits (up-up), the device continues to scatter light only in the forward direction since its symmetry is not modified by the modulation. Accordingly, we observe side-bands mostly scattered forward and almost fully suppressed in back scattering (dashed circles in panels a and c). However, if the modulation has a phase difference of $\alpha=\pi$, the situation is reversed. This is illustrated by the blue arrows inside the second qubit (up-down) in the insets of Fig.~\ref{fig3}b and d, corresponding to an additional phase factor of $-1$. While the inelastic backscattering is now highly likely as shown in Fig.~\ref{fig3}d, the side-bands scattered forward from the first qubit destructively interfere with the ones scattered from the second one due to the additional phase shift and thus preventing forward scattering as shown in Fig.~\ref{fig3}b.

%%%%%%%%%%%%%%%%%%%%%%%%%%%%%%%%%%%%%%%%%%%%%%%%%%%%%%%%%%%
%%%%%%%%%%%%%%%%%%%%%%%%%%%%%%%%%%%%%%%%%%%%%%%%%%%%%%%%%%%%

\begin{figure}[t]
\centering
\includegraphics[width=\columnwidth]{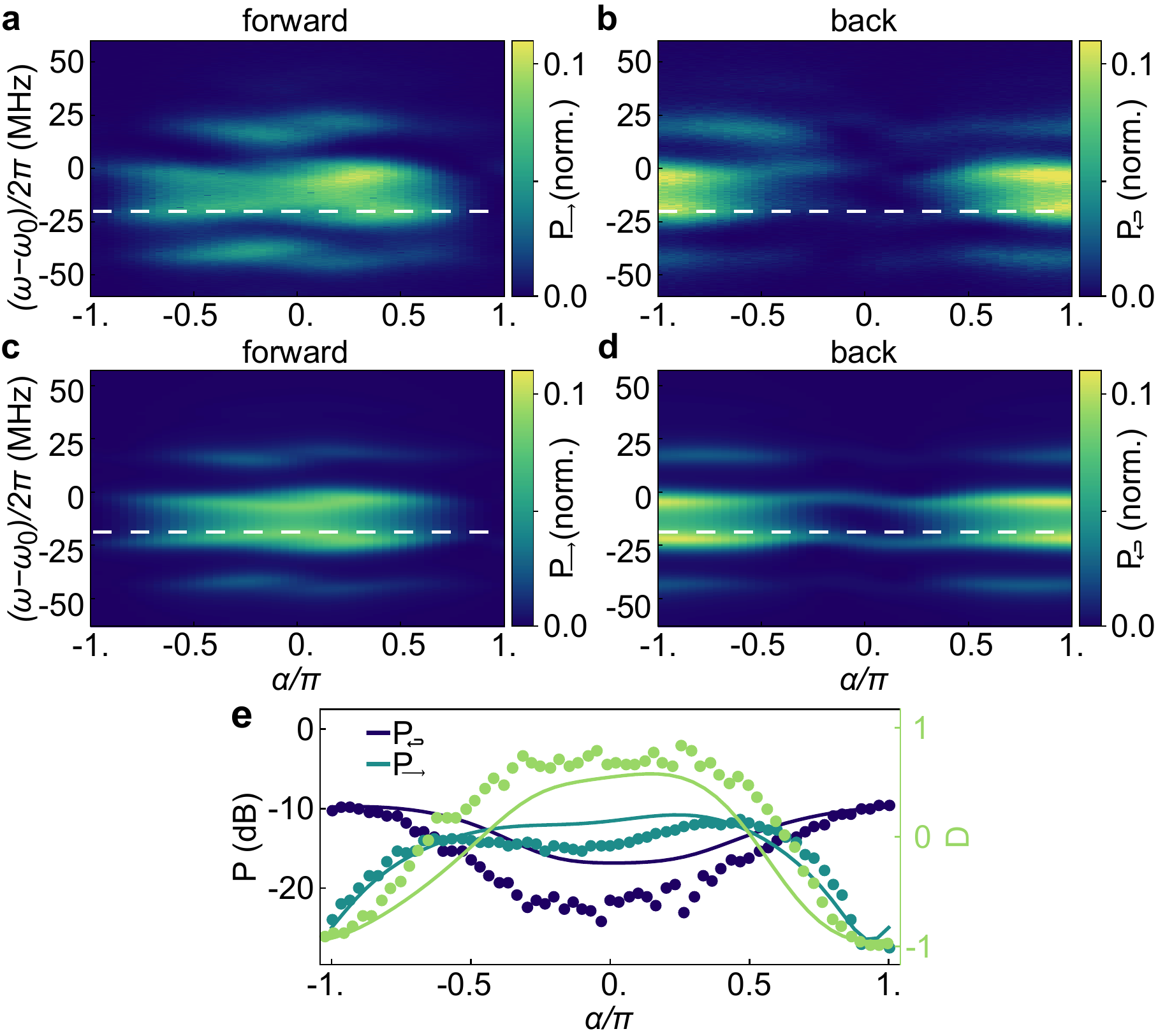}
\caption{\textbf{Coherent inelastic scattering spectrum of the Stokes component.}
\textbf{a(b)}, Measured and normalized Stokes power as a function of probe frequency detuning $\omega-\omega_0$ and relative phase between modulation tones $\alpha$ measured in transmission (reflection) at a fixed modulation amplitude $A_m/ (2\pi) = 30$\,MHz and modulation frequency $\Omega/ (2\pi) = 20$\,MHz. \textbf{c(d)}, Theoretically predicted transmission (reflection) spectrum on the same scale. \textbf{e}, Measured coherent inelastic scattering as a function of $\alpha$ at $(\omega - \omega_0)/ (2\pi) = \Omega/ (2\pi) = -20$ MHz (points) and theory (solid lines). Scattering directivity $D$ is shown in light green.} \label{fig4}
\end{figure}

In order to better illustrate the phase and detuning dependence of the interference conditions 
%with the change of the modulation phase difference $\alpha$, 
we extract the coherent scattering power of the Stokes component over the full range of $\alpha$ and for finite detuning from the qubit resonances at $\omega_0$. For this measurement the detection frequency is always detuned by the chosen modulation frequency $\Omega/(2\pi)=20$\,MHz from the probe tone at frequency $\omega$. 
%which is swept. 
Here we detect both the transmitted and reflected scattered Stokes light with the two channels of the digitizer simultaneously for $A_m/(2\pi)=30$\,MHz. The obtained intensity in transmission and reflection is shown in Fig.~\ref{fig4}a and b. 
%and the corresponding theory in Fig.~\ref{fig4}c and d.
%and its overall dependence on $\alpha$ is rather intricate. 
We observe resonances at probe frequencies $\omega_0$, $\omega_0\pm \Omega$, and $\omega_0-2\Omega$ and their overall dependence on $\alpha$ is clearly pronounced and opposite in sign for forward and backward scattering. 
These experimental results are in very good agreement with the theoretical model shown in Fig.~\ref{fig4}(b,d), see Methods for details.
%Dependence of the scattering parameters on the phase difference $\alpha$ as well as the directivity in logarithmic scale [$D = 10\log_{10}(P_{\hookleftarrow}/P_{\rightarrow})$], shown in Fig. \ref{fig4}e, demonstrate the transition between different regimes when light is mostly scattered back ($D < 0$), forward ($D > 0$), or symmetrically in both directions ($D = 0$). For the relative phase $\alpha/\pi = 0$, experimentally measured suppression of backscattered signal reaches $20.6$ dB, whereas for $\alpha/\pi = \pm 1$, where light scattered forward interfere destructively, we suppress up to $14.6$ dB of transmitted light.
%%%%%%%%% DIODE DESCRIPTION %%%%%%%%%

The measured dependence of the scattering parameters on the phase difference $\alpha$ as well as the directivity $D = (P_{\rightarrow}-P_{\hookleftarrow})/(P_{\rightarrow}+P_{\hookleftarrow})$ is shown in Fig.~\ref{fig4}e for $(\omega-\omega_0)/(2\pi)=\Omega/(2\pi) = - 20$\,MHz (dashed white lines in panels a-d) together with theory. This shows phase selective control to enter the regimes when light is mostly scattered back ($D < 0$), forward ($D > 0$), or symmetrically in both directions ($D = 0$). The measured directivity demonstrates high diode efficiency that can be set continuously between 0.84 and -0.99.
%%%%%%%%%%%%%%%%%%%%%%%%%%%%%%%%%%%%%%%%%%%%%%%%%%%%%%%%%%%
%%%%%%%%%%%%%%%%%%%%%%%%%%%%%%%%%%%%%%%%%%%%%%%%%%%%%%%%%%%%

\vspace{0.25cm}
\noindent
\section*{Discussion}
Besides the demonstrated high scattering directionality, at the relative phase $\alpha/\pi = \pm 1$, our system exhibits a gyrator-like behavior and  transmits a signal unchanged in one direction whereas the reverse traveling signal experiences a $\pi$ phase-shift. For $\alpha/\pi = \pm 0.3$, our device demonstrates isolator properties, which might be enhanced by further tuning the device or by extending the principle to a larger number of qubits, details can be found in the Supplementary Discussion. This nonreciprocity relies on the qubits working in the linear regime, which would naturally limit the operation of the device to low powers $(\Omega_{R}/\gammarad)^2  \ll 1$. In contrast, our calculations  indicate that  scattering remains directional up to intermediate drive powers 
%$(\Omega_{R}/\gammarad)^2  \sim 1$. The theoretically estimated upper limit is 
$(\Omega_{R}/\gammarad)^2 \lesssim 9$ beyond which the inelastic scattering is fully suppressed, see Supplementary Figures.

We have demonstrated 
%a directional quantum device
an on-chip microwave photon router that can be switched \textit{in-situ} between scattering photons back, forward, or symmetrically in both directions.
% a system is perfect for the on-chip microwave light routing. 
While it is limited in dynamic range, it is fully compatible with modern superconducting quantum computing devices \cite{oliver2020b,Reuer2022} that operate in the single and few photon regime. 
%Although the bandwidth of an inelastically scattered signal can be modified only by changing the qubit-waveguide coupling strength, 
The suppression strength can be modified by the modulation amplitude and the signal frequency can be shifted and fine-tuned \textit{in-situ} by changing the modulation frequency. A larger range of frequency bands can be accessed by working at 
% While our device was designed to perform at the $\lambda/4$ effective distance between the qubits,  the same approach works for any 
odd multiples of the $\lambda/4$ boundary condition. 
%This enables future applications with an even wider frequency tunability range. 
%Finally, the 
In the future, this device principle could be extended to multiple nodes, see Methods and Supplementary Information, in order to realize topologically protected states \cite{lu2016topological}, as a part of a hardware implementation of Gottesman-Kitaev-Preskill codes \cite{rymarz2021hardware}, or to route microwave radiation for the realization of chiral networks \cite{lodahl2017chiral}.

\vspace{0.25cm}
\noindent
\normalsize
\section*{Methods}
\footnotesize
\textbf{Calculation of scattering spectra.}
In this section we present the general approach to calculate photon scattering from an array of qubits with  time-modulated resonance frequencies. Such a device is characterized by the following effective non-Hermitian Hamiltonian ~\cite{Caneva2015}:
\begin{multline}\label{eq:H}
    H(t)=\sum\limits_{j}[\omega_0^{(j)}(t)-\rmi \Gamma_2^{(j)}]\sigma_j^\dag \sigma_j-\frac{\rmi\Gamma_1}2 \sum\limits_{j,k}\e^{\rmi \varphi|j-k|}\sigma_j^\dag \sigma_k
    \\+\frac{\Omega_{\rm R}}{2}\sum\limits_j (\sigma_j^\dag\e^{\rmi\varphi j-\rmi \omega t}+{\rm H.c.})\:.
\end{multline}
Here $\sigma_j$ are the raising operators, $\Gamma_1$ is the (radiative) relaxation rate between the $|1\rangle$ and $|0\rangle$ qubit states, $\Gamma_2^{(j)}$ is the decay rate of the coherence between the $|1\rangle$ and $|0\rangle$ states, $\varphi=\omega_0d/c$ is the phase gained by light traveling between the qubits with propagation velocity $c$. The Rabi frequency $\Omega_R$ quantifies the incident wave amplitude and 
\begin{equation}
    \omega_0^{(j)}(t)=\omega_0+A_m\cos(\Omega t+\alpha_j)
\end{equation}
are the time-dependent qubit resonance frequencies.  %All the decay and dephasing of the qubits besides emission into the waveguide is incorporated into the rates $\gamma_j$.
The Hamiltonian Eq.~\eqref{eq:H} assumes the usual rotating wave and Markovian approximations.
Here, we are interested in the case of weak coherent driving
%\hl{In this case only the single-excited states of the qubit array are populated. }The 
where the wavefunction can be approximately written as 
\begin{equation}
\psi=|0\rangle+\sum\limits_j p_j \sigma_j^\dag |0\rangle\:.
\end{equation}
The amplitudes $p_j$ describe the coherence between the ground and excited states and can be found from the following Schr\"odinger equation:
\begin{multline}
\rmi\frac{\rmd }{\rmd t}p_j(t)=[\omega_0^{(j)}(t)-\rmi\Gamma_2^{(j)}]p_j\\-\frac{\rmi\Gamma_1}2 \sum\limits_{k}\e^{\rmi \varphi|j-k|}p_k+\frac{\Omega_{\rm R}}{2}\,\e^{\rmi\varphi j-\rmi \omega t} \:.
\end{multline}
It is convenient to seek the solution in the form
\begin{equation}
p_j(t)=\sum\limits_{n=-\infty}^\infty p_j^{(n)}\e^{-\rmi (\omega+n\Omega)t}\:,
\end{equation}
so that the amplitudes 
$p_j^{(n)}$ are determined by the linear system \cite{Poddubny2021Ratchet}
\begin{multline}\label{eq:pjm}
(\omega+n\Omega)p_j^{(n)}=(\omega_0-\rmi \Gamma_2^{(j)})p_j^{(n)}
+\frac{A_m}{2}(\e^{\rmi \alpha_j}p_j^{(n-1)}+\e^{\rmi \alpha_j}p_j^{(n+1)})
\\-\frac{\rmi\Gamma_1}2 \sum\limits_{k}\e^{\rmi \varphi|j-k|}p_k^{(n)}+\frac{\Omega_{\rm R}}{2}\, \e^{\rmi\varphi j}\delta_{m,0}\:.
\end{multline}
After the amplitudes $p_j^{(n)}$ have been found numerically, we calculate the coefficients  $r^{(n)}$ and $t^{(n)}$ 
\begin{align}
r^{(n)}&=-\frac{\rmi\Gamma_1}{\Omega_{\rm R}}\sum\limits_j \e^{\rmi\varphi j}p_j^{(n)},\quad \\ t^{(n)}&=\delta_{n,0}-\frac{\rmi\Gamma_1}{\Omega_{\rm R}}\sum\limits_j \e^{-\rmi\varphi j}p_j^{(n)}\:,
\end{align}
that describe the backward (forward) scattering process with the frequency change $\omega \to \omega+n\Omega$. In the general case, the system of equations 
\eqref{eq:pjm} is to be solved numerically. However,  it is possible to obtain an analytical solution in the particular case of a single qubit~\cite{Marquardt2006}. In this case we find
\begin{equation}\label{eq:pnanswer}
p^{(n)}=\frac{\Omega_{\rm R}}{2}\,
\sum_{n'=-\infty}^\infty
\frac{ J_{n'-n}(A_m/\Omega)J_{n'}(A_m/\Omega)}{\omega+n'\Omega-\omega_0+\rmi\Gamma_2}\:.
\end{equation}
For elastic scattering ($n=0$) Eq.~\eqref{eq:pnanswer} leads to Eq.~\eqref{coh_oneQ} in the main text.

\noindent
\textbf{Resonance fluorescence of the time-modulated device.}
Here we describe the procedure to calculate the nested Mollow triplets shown in Fig.~\ref{fig2}. 
%in the spectrum of the incoherent resonance fluorescence  of a single qubit with time-modulated resonance frequency. 
The state of the qubit can be conveniently represented as vector $\bm S$ of the spin 1/2, where $|1\rangle$ and $|0\rangle$ states  correspond to $S_z=1/2$ and $-1/2$, respectively. The dynamics $\bm S(t)$ is governed by the Bloch equation that reads
\begin{equation}\label{eq:dS}
\frac{\rmd\bm S}{\rmd t}=\bm S\times \widetilde{\bm \Omega}(t)-\bm\Gamma (\bm S - \bm S_0)
\end{equation}
where  
\begin{align}
 \widetilde{\bm \Omega}(t) = [\Omega_R \cos\omega t, \Omega_R \sin\omega t, \omega_0+\Delta\omega\cos(\Omega t+\alpha)]\,
\end{align}
is the time-dependent effective magnetic field,
$\bm S_0 = [0,0,-1/2]$ is the equilibrium spin, and the spin relaxation term reads $\bm\Gamma (\bm S - \bm S_0)\equiv [\Gamma_2 S_x,\Gamma_2 S_y,\Gamma_1 (S_z+1/2)] $. 
The emission spectrum is determined by the spin correlation function
\begin{equation}\label{eq:Is}
I(\omega)\propto {\rm Re}\!\int_0^\infty \rmd t\:\e^{-\rmi \omega \tau}\langle\!\langle S_+(t+\tau)S_-(t)\rangle\!\rangle_t, 
\end{equation}
where $S_\pm=S_x\pm \rmi S_y$ and the double angular brackets denote averaging over the absolute time $t$. Equation~\eqref{eq:Is} establishes the correspondence between the emission spectrum in the considered quantum optics problem and the electron spin noise spectrum in the conditions of electron paramagnetic resonance, when the electron is subject to two magnetic fields, a constant one and an oscillating one~\cite{poshakinskiy2020spin}.

In the theory of magnetic resonance, the standard trick to solve Eq.~\eqref{eq:dS} analytically is to switch to a reference frame rotating around the $z$ axis with the drive frequency $\omega$. There, the spin dynamics is governed by the same Eq.~\eqref{eq:dS} but $ \widetilde{\bm \Omega} (t)$ shall be replaced with
\begin{align}
\bm \Omega'(t) = [\Omega_R , 0, \omega_0-\omega+\Delta\omega\cos\Omega t]\,.
\end{align}
In the absence of modulation, $\Delta\omega = 0$, the effective magnetic field $ \bm \Omega'(t)$ would be constant and its amplitude
\begin{align}
\Omega_R' = \sqrt{\Omega_R^2 +(\omega_0-\omega)^2}
\end{align}
would determine the splitting in the conventional Mollow triplet. 

The presence of modulation can be accounted for by repeating the trick and switching to yet another frame rotating with frequency $\Omega_R'$ with respect to the previous one. There, $\bm\Omega' (t)$ is replaced with
\begin{align}
\bm \Omega'' =\frac{\Omega_R \Delta\omega}{2{\Omega_R'}^2} [\omega-\omega_0 , 0, \Omega_R]+ \Big(1-\frac{\Omega}{\Omega_R'}\Big)[\Omega_R , 0, \omega_0-\omega]\,,
\end{align}
where we neglected all oscillating terms, since they average to zero. The amplitude of $\bm \Omega''$ determines the splitting of the nested Mollow triplet
\begin{align}
 \Omega_R'' =\sqrt{\left(\frac{\Omega_R \Delta\omega}{2 \Omega_R'}\right)^2 + (\Omega_R' - \Omega)^2}\,.
\end{align}
 Returning back to the initial reference frame, we obtain nine possible emission frequencies
\begin{align}
\omega_{p,q} = \omega + p \Omega_R' + q \Omega_R''\,,
\end{align}
where $p,q = 0,\pm 1$ enumerate the components of the two nested Mollow triplets. In the above analytical solution, we used twice the rotating wave approximation, which  is valid provided $\Delta\omega \ll \Omega_R \ll \omega_0$. 

% large Rabi frequencies when the electromagnetic field can be considered semiclassically.

%In such case the considered quantum optics problem is equivalent to the problem of the calculation  of spin noise spectrum in the theory of electron paramagnetic resonance for a spin 1/2 subjected to constant magnetic field and an oscillating radiofrequency field~\cite{poshakinskiy2020spin}. The role of the constant magnetic field is played by the Rabi frequency $\omega_R$ and the radiofrequency field in our case is replaced by the time-dependent resonance frequency modulation. Specifically, the evolution of the spin components is described by the equation

\noindent
\textbf{Data normalization.}
We normalize the transmission spectra $|t_0|^2$ shown in Fig.~\ref{fig1}d, e and Fig.~\ref{fig2}a, b by dividing the background transmission coefficient $|t_0|^2 = |t|^2/|t_{bg}|^2$. Here, $|t_{bg}|^2$ is measured with both qubits tuned out of the frequency range of interest and $|t|^2$ is measured with the qubit tuned to the desired frequency. This method normalizes the gain in the system and compensates for the frequency-dependent transmission properties of the drive and detection lines.

The power spectral density of the measured resonance fluorescence spectrum $S(\omega)$ shown in Fig.~\ref{fig2}c, as well as the coherent inelastic scattering spectra shown in Fig.~\ref{fig4}a, b were scaled to the numerically predicted value. The latter relies on the qubit parameters extracted from the transmission measurements, the chosen modulation frequency and the independently calibrated modulation amplitude.

\vspace{0.25cm}
\noindent
\normalsize
\textbf{Data availability}\\
\footnotesize
All datasets and analysis files used in this study will be made available on the Zenodo repository before publication. 

\vspace{0.25cm}
\noindent
\normalsize
\textbf{Acknowledgments}\\
\footnotesize
The authors thank L. Drmic and P. Zielinski for software development, and the MIBA workshop and the IST nanofabrication facility for technical support. This work was supported by the Austrian Science Fund (FWF) through BeyondC (F7105) and IST Austria. E.R. is the recipient of a DOC fellowship of the Austrian Academy of Sciences at IST Austria. J.M.F. and M.Z. acknowledge support from the European Research Council under grant agreement No 758053 (ERC StG QUNNECT) and a NOMIS foundation research grant. The work of A.N.P. and A.V.P. has been supported by the Russian Science Foundation under the grant No 20-12-00194.

\vspace{0.25cm}
\noindent
\normalsize
\textbf{Author contributions} \\
\footnotesize
E.S.R. designed and fabricated the samples, worked on the setup and performed the measurements. A.V.P. and A.N.P. developed the theory and together with E.S.R conducted the data analysis. M.Z. contributed to building the measurement setup and R.S. to the resonance fluorescence data acquisition code. E.S.R. and A.N.P. wrote the manuscript with contributions from all authors. J.M.F. supervised this work.

\vspace{0.25cm}
\noindent
\normalsize
\textbf{Competing interests} \\
\footnotesize
The authors declare no competing interests.

\renewcommand*{\bibfont}{\footnotesize}
\bibliography{bibliography}

\begin{thebibliography}{44}%
\makeatletter
\providecommand \@ifxundefined [1]{%
 \@ifx{#1\undefined}
}%
\providecommand \@ifnum [1]{%
 \ifnum #1\expandafter \@firstoftwo
 \else \expandafter \@secondoftwo
 \fi
}%
\providecommand \@ifx [1]{%
 \ifx #1\expandafter \@firstoftwo
 \else \expandafter \@secondoftwo
 \fi
}%
\providecommand \natexlab [1]{#1}%
\providecommand \enquote  [1]{``#1''}%
\providecommand \bibnamefont  [1]{#1}%
\providecommand \bibfnamefont [1]{#1}%
\providecommand \citenamefont [1]{#1}%
\providecommand \href@noop [0]{\@secondoftwo}%
\providecommand \href [0]{\begingroup \@sanitize@url \@href}%
\providecommand \@href[1]{\@@startlink{#1}\@@href}%
\providecommand \@@href[1]{\endgroup#1\@@endlink}%
\providecommand \@sanitize@url [0]{\catcode `\\12\catcode `\$12\catcode
  `\&12\catcode `\#12\catcode `\^12\catcode `\_12\catcode `\%12\relax}%
\providecommand \@@startlink[1]{}%
\providecommand \@@endlink[0]{}%
\providecommand \url  [0]{\begingroup\@sanitize@url \@url }%
\providecommand \@url [1]{\endgroup\@href {#1}{\urlprefix }}%
\providecommand \urlprefix  [0]{URL }%
\providecommand \Eprint [0]{\href }%
\providecommand \doibase [0]{https://doi.org/}%
\providecommand \selectlanguage [0]{\@gobble}%
\providecommand \bibinfo  [0]{\@secondoftwo}%
\providecommand \bibfield  [0]{\@secondoftwo}%
\providecommand \translation [1]{[#1]}%
\providecommand \BibitemOpen [0]{}%
\providecommand \bibitemStop [0]{}%
\providecommand \bibitemNoStop [0]{.\EOS\space}%
\providecommand \EOS [0]{\spacefactor3000\relax}%
\providecommand \BibitemShut  [1]{\csname bibitem#1\endcsname}%
\let\auto@bib@innerbib\@empty
%</preamble>
\bibitem [{\citenamefont {S\"{o}llner}\ \emph {et~al.}(2015)\citenamefont
  {S\"{o}llner}, \citenamefont {Mahmoodian}, \citenamefont {Hansen},
  \citenamefont {Midolo}, \citenamefont {Javadi}, \citenamefont
  {Kir{\v{s}}ansk{\.{e}}}, \citenamefont {Pregnolato}, \citenamefont {El-Ella},
  \citenamefont {Lee}, \citenamefont {Song}, \citenamefont {Stobbe},\ and\
  \citenamefont {Lodahl}}]{Sllner2015}%
  \BibitemOpen
  \bibfield  {author} {\bibinfo {author} {\bibfnamefont {I.}~\bibnamefont
  {S\"{o}llner}}, \bibinfo {author} {\bibfnamefont {S.}~\bibnamefont
  {Mahmoodian}}, \bibinfo {author} {\bibfnamefont {S.~L.}\ \bibnamefont
  {Hansen}}, \bibinfo {author} {\bibfnamefont {L.}~\bibnamefont {Midolo}},
  \bibinfo {author} {\bibfnamefont {A.}~\bibnamefont {Javadi}}, \bibinfo
  {author} {\bibfnamefont {G.}~\bibnamefont {Kir{\v{s}}ansk{\.{e}}}}, \bibinfo
  {author} {\bibfnamefont {T.}~\bibnamefont {Pregnolato}}, \bibinfo {author}
  {\bibfnamefont {H.}~\bibnamefont {El-Ella}}, \bibinfo {author} {\bibfnamefont
  {E.~H.}\ \bibnamefont {Lee}}, \bibinfo {author} {\bibfnamefont {J.~D.}\
  \bibnamefont {Song}}, \bibinfo {author} {\bibfnamefont {S.}~\bibnamefont
  {Stobbe}},\ and\ \bibinfo {author} {\bibfnamefont {P.}~\bibnamefont
  {Lodahl}},\ }\bibfield  {title} {\bibinfo {title} {Deterministic
  photon{\textendash}emitter coupling in chiral photonic circuits},\ }\href
  {https://doi.org/10.1038/nnano.2015.159} {\bibfield  {journal} {\bibinfo
  {journal} {Nature Nanotech}\ }\textbf {\bibinfo {volume} {10}},\ \bibinfo
  {pages} {775} (\bibinfo {year} {2015})}\BibitemShut {NoStop}%
\bibitem [{\citenamefont {Spitzer}\ \emph {et~al.}(2018)\citenamefont
  {Spitzer}, \citenamefont {Poddubny}, \citenamefont {Akimov}, \citenamefont
  {Sapega}, \citenamefont {Klompmaker}, \citenamefont {Kreilkamp},
  \citenamefont {Litvin}, \citenamefont {Jede}, \citenamefont {Karczewski},
  \citenamefont {Wiater}, \citenamefont {Wojtowicz}, \citenamefont {Yakovlev},\
  and\ \citenamefont {Bayer}}]{Spitzer2018}%
  \BibitemOpen
  \bibfield  {author} {\bibinfo {author} {\bibfnamefont {F.}~\bibnamefont
  {Spitzer}}, \bibinfo {author} {\bibfnamefont {A.~N.}\ \bibnamefont
  {Poddubny}}, \bibinfo {author} {\bibfnamefont {I.~A.}\ \bibnamefont
  {Akimov}}, \bibinfo {author} {\bibfnamefont {V.~F.}\ \bibnamefont {Sapega}},
  \bibinfo {author} {\bibfnamefont {L.}~\bibnamefont {Klompmaker}}, \bibinfo
  {author} {\bibfnamefont {L.~E.}\ \bibnamefont {Kreilkamp}}, \bibinfo {author}
  {\bibfnamefont {L.~V.}\ \bibnamefont {Litvin}}, \bibinfo {author}
  {\bibfnamefont {R.}~\bibnamefont {Jede}}, \bibinfo {author} {\bibfnamefont
  {G.}~\bibnamefont {Karczewski}}, \bibinfo {author} {\bibfnamefont
  {M.}~\bibnamefont {Wiater}}, \bibinfo {author} {\bibfnamefont
  {T.}~\bibnamefont {Wojtowicz}}, \bibinfo {author} {\bibfnamefont {D.~R.}\
  \bibnamefont {Yakovlev}},\ and\ \bibinfo {author} {\bibfnamefont
  {M.}~\bibnamefont {Bayer}},\ }\bibfield  {title} {\bibinfo {title} {Routing
  the emission of a near-surface light source by a magnetic field},\ }\href
  {https://doi.org/10.1038/s41567-018-0232-7} {\bibfield  {journal} {\bibinfo
  {journal} {Nature Physics}\ }\textbf {\bibinfo {volume} {14}},\ \bibinfo
  {pages} {1043} (\bibinfo {year} {2018})}\BibitemShut {NoStop}%
\bibitem [{\citenamefont {Rosario~Hamann}\ \emph {et~al.}(2018)\citenamefont
  {Rosario~Hamann}, \citenamefont {M\"uller}, \citenamefont {Jerger},
  \citenamefont {Zanner}, \citenamefont {Combes}, \citenamefont {Pletyukhov},
  \citenamefont {Weides}, \citenamefont {Stace},\ and\ \citenamefont
  {Fedorov}}]{hamann2018nonreciprocity}%
  \BibitemOpen
  \bibfield  {author} {\bibinfo {author} {\bibfnamefont {A.}~\bibnamefont
  {Rosario~Hamann}}, \bibinfo {author} {\bibfnamefont {C.}~\bibnamefont
  {M\"uller}}, \bibinfo {author} {\bibfnamefont {M.}~\bibnamefont {Jerger}},
  \bibinfo {author} {\bibfnamefont {M.}~\bibnamefont {Zanner}}, \bibinfo
  {author} {\bibfnamefont {J.}~\bibnamefont {Combes}}, \bibinfo {author}
  {\bibfnamefont {M.}~\bibnamefont {Pletyukhov}}, \bibinfo {author}
  {\bibfnamefont {M.}~\bibnamefont {Weides}}, \bibinfo {author} {\bibfnamefont
  {T.~M.}\ \bibnamefont {Stace}},\ and\ \bibinfo {author} {\bibfnamefont
  {A.}~\bibnamefont {Fedorov}},\ }\bibfield  {title} {\bibinfo {title}
  {Nonreciprocity realized with quantum nonlinearity},\ }\href
  {https://doi.org/10.1103/PhysRevLett.121.123601} {\bibfield  {journal}
  {\bibinfo  {journal} {Phys. Rev. Lett.}\ }\textbf {\bibinfo {volume} {121}},\
  \bibinfo {pages} {123601} (\bibinfo {year} {2018})}\BibitemShut {NoStop}%
\bibitem [{\citenamefont {Poulton}\ \emph {et~al.}(2012)\citenamefont
  {Poulton}, \citenamefont {Pant}, \citenamefont {Byrnes}, \citenamefont {Fan},
  \citenamefont {Steel},\ and\ \citenamefont {Eggleton}}]{poulton2012design}%
  \BibitemOpen
  \bibfield  {author} {\bibinfo {author} {\bibfnamefont {C.~G.}\ \bibnamefont
  {Poulton}}, \bibinfo {author} {\bibfnamefont {R.}~\bibnamefont {Pant}},
  \bibinfo {author} {\bibfnamefont {A.}~\bibnamefont {Byrnes}}, \bibinfo
  {author} {\bibfnamefont {S.}~\bibnamefont {Fan}}, \bibinfo {author}
  {\bibfnamefont {M.~J.}\ \bibnamefont {Steel}},\ and\ \bibinfo {author}
  {\bibfnamefont {B.~J.}\ \bibnamefont {Eggleton}},\ }\bibfield  {title}
  {\bibinfo {title} {Design for broadband on-chip isolator using stimulated
  brillouin scattering in dispersion-engineered chalcogenide waveguides},\
  }\href {https://doi.org/10.1364/OE.20.021235} {\bibfield  {journal} {\bibinfo
   {journal} {Opt. Express}\ }\textbf {\bibinfo {volume} {20}},\ \bibinfo
  {pages} {21235} (\bibinfo {year} {2012})}\BibitemShut {NoStop}%
\bibitem [{\citenamefont {Kittlaus}\ \emph {et~al.}(2017)\citenamefont
  {Kittlaus}, \citenamefont {Otterstrom},\ and\ \citenamefont
  {Rakich}}]{kittlaus2017chip}%
  \BibitemOpen
  \bibfield  {author} {\bibinfo {author} {\bibfnamefont {E.~A.}\ \bibnamefont
  {Kittlaus}}, \bibinfo {author} {\bibfnamefont {N.~T.}\ \bibnamefont
  {Otterstrom}},\ and\ \bibinfo {author} {\bibfnamefont {P.~T.}\ \bibnamefont
  {Rakich}},\ }\bibfield  {title} {\bibinfo {title} {On-chip inter-modal
  {B}rillouin scattering},\ }\href {https://doi.org/10.1038/ncomms15819}
  {\bibfield  {journal} {\bibinfo  {journal} {Nature communications}\ }\textbf
  {\bibinfo {volume} {8}},\ \bibinfo {pages} {1} (\bibinfo {year}
  {2017})}\BibitemShut {NoStop}%
\bibitem [{\citenamefont {Safavi-Naeini}\ \emph {et~al.}(2019)\citenamefont
  {Safavi-Naeini}, \citenamefont {Van~Thourhout}, \citenamefont {Baets},\ and\
  \citenamefont {Van~Laer}}]{safavi2019controlling}%
  \BibitemOpen
  \bibfield  {author} {\bibinfo {author} {\bibfnamefont {A.~H.}\ \bibnamefont
  {Safavi-Naeini}}, \bibinfo {author} {\bibfnamefont {D.}~\bibnamefont
  {Van~Thourhout}}, \bibinfo {author} {\bibfnamefont {R.}~\bibnamefont
  {Baets}},\ and\ \bibinfo {author} {\bibfnamefont {R.}~\bibnamefont
  {Van~Laer}},\ }\bibfield  {title} {\bibinfo {title} {Controlling phonons and
  photons at the wavelength scale: integrated photonics meets integrated
  phononics},\ }\href {https://doi.org/10.1364/OPTICA.6.000213} {\bibfield
  {journal} {\bibinfo  {journal} {Optica}\ }\textbf {\bibinfo {volume} {6}},\
  \bibinfo {pages} {213} (\bibinfo {year} {2019})}\BibitemShut {NoStop}%
\bibitem [{\citenamefont {Reuer}\ \emph {et~al.}(2022)\citenamefont {Reuer},
  \citenamefont {Besse}, \citenamefont {Wernli}, \citenamefont {Magnard},
  \citenamefont {Kurpiers}, \citenamefont {Norris}, \citenamefont {Wallraff},\
  and\ \citenamefont {Eichler}}]{Reuer2022}%
  \BibitemOpen
  \bibfield  {author} {\bibinfo {author} {\bibfnamefont {K.}~\bibnamefont
  {Reuer}}, \bibinfo {author} {\bibfnamefont {J.-C.}\ \bibnamefont {Besse}},
  \bibinfo {author} {\bibfnamefont {L.}~\bibnamefont {Wernli}}, \bibinfo
  {author} {\bibfnamefont {P.}~\bibnamefont {Magnard}}, \bibinfo {author}
  {\bibfnamefont {P.}~\bibnamefont {Kurpiers}}, \bibinfo {author}
  {\bibfnamefont {G.~J.}\ \bibnamefont {Norris}}, \bibinfo {author}
  {\bibfnamefont {A.}~\bibnamefont {Wallraff}},\ and\ \bibinfo {author}
  {\bibfnamefont {C.}~\bibnamefont {Eichler}},\ }\bibfield  {title} {\bibinfo
  {title} {Realization of a universal quantum gate set for itinerant microwave
  photons},\ }\href {https://doi.org/10.1103/PhysRevX.12.011008} {\bibfield
  {journal} {\bibinfo  {journal} {Phys. Rev. X}\ }\textbf {\bibinfo {volume}
  {12}},\ \bibinfo {pages} {011008} (\bibinfo {year} {2022})}\BibitemShut
  {NoStop}%
\bibitem [{\citenamefont {Astafiev}\ \emph {et~al.}(2010)\citenamefont
  {Astafiev}, \citenamefont {Zagoskin}, \citenamefont {Abdumalikov},
  \citenamefont {Pashkin}, \citenamefont {Yamamoto}, \citenamefont {Inomata},
  \citenamefont {Nakamura},\ and\ \citenamefont
  {Tsai}}]{astafiev2010resonance}%
  \BibitemOpen
  \bibfield  {author} {\bibinfo {author} {\bibfnamefont {O.}~\bibnamefont
  {Astafiev}}, \bibinfo {author} {\bibfnamefont {A.~M.}\ \bibnamefont
  {Zagoskin}}, \bibinfo {author} {\bibfnamefont {A.}~\bibnamefont
  {Abdumalikov}}, \bibinfo {author} {\bibfnamefont {Y.~A.}\ \bibnamefont
  {Pashkin}}, \bibinfo {author} {\bibfnamefont {T.}~\bibnamefont {Yamamoto}},
  \bibinfo {author} {\bibfnamefont {K.}~\bibnamefont {Inomata}}, \bibinfo
  {author} {\bibfnamefont {Y.}~\bibnamefont {Nakamura}},\ and\ \bibinfo
  {author} {\bibfnamefont {J.~S.}\ \bibnamefont {Tsai}},\ }\bibfield  {title}
  {\bibinfo {title} {Resonance fluorescence of a single artificial atom},\
  }\href {https://doi.org/10.1126/science.1181918} {\bibfield  {journal}
  {\bibinfo  {journal} {Science}\ }\textbf {\bibinfo {volume} {327}},\ \bibinfo
  {pages} {840} (\bibinfo {year} {2010})}\BibitemShut {NoStop}%
\bibitem [{\citenamefont {van Loo}\ \emph {et~al.}(2013)\citenamefont {van
  Loo}, \citenamefont {Fedorov}, \citenamefont {Lalumi{\`e}re}, \citenamefont
  {Sanders}, \citenamefont {Blais},\ and\ \citenamefont
  {Wallraff}}]{vanLoo1494}%
  \BibitemOpen
  \bibfield  {author} {\bibinfo {author} {\bibfnamefont {A.~F.}\ \bibnamefont
  {van Loo}}, \bibinfo {author} {\bibfnamefont {A.}~\bibnamefont {Fedorov}},
  \bibinfo {author} {\bibfnamefont {K.}~\bibnamefont {Lalumi{\`e}re}}, \bibinfo
  {author} {\bibfnamefont {B.~C.}\ \bibnamefont {Sanders}}, \bibinfo {author}
  {\bibfnamefont {A.}~\bibnamefont {Blais}},\ and\ \bibinfo {author}
  {\bibfnamefont {A.}~\bibnamefont {Wallraff}},\ }\bibfield  {title} {\bibinfo
  {title} {Photon-mediated interactions between distant artificial atoms},\
  }\href {https://doi.org/10.1126/science.1244324} {\bibfield  {journal}
  {\bibinfo  {journal} {Science}\ }\textbf {\bibinfo {volume} {342}},\ \bibinfo
  {pages} {1494} (\bibinfo {year} {2013})}\BibitemShut {NoStop}%
\bibitem [{\citenamefont {Kannan}\ \emph {et~al.}(2022)\citenamefont {Kannan},
  \citenamefont {Almanakly}, \citenamefont {Sung}, \citenamefont {Di~Paolo},
  \citenamefont {Rower}, \citenamefont {Braum\"uller}, \citenamefont
  {Melville}, \citenamefont {Niedzielski}, \citenamefont {Karamlou},
  \citenamefont {Serniak}, \citenamefont {Veps\"al\"ainen}, \citenamefont
  {Schwartz}, \citenamefont {Yoder}, \citenamefont {Winik}, \citenamefont
  {Wang}, \citenamefont {Orlando}, \citenamefont {Gustavsson}, \citenamefont
  {Grover},\ and\ \citenamefont {Oliver}}]{Kannan2022}%
  \BibitemOpen
  \bibfield  {author} {\bibinfo {author} {\bibfnamefont {B.}~\bibnamefont
  {Kannan}}, \bibinfo {author} {\bibfnamefont {A.}~\bibnamefont {Almanakly}},
  \bibinfo {author} {\bibfnamefont {Y.}~\bibnamefont {Sung}}, \bibinfo {author}
  {\bibfnamefont {A.}~\bibnamefont {Di~Paolo}}, \bibinfo {author}
  {\bibfnamefont {D.~A.}\ \bibnamefont {Rower}}, \bibinfo {author}
  {\bibfnamefont {J.}~\bibnamefont {Braum\"uller}}, \bibinfo {author}
  {\bibfnamefont {A.}~\bibnamefont {Melville}}, \bibinfo {author}
  {\bibfnamefont {B.~M.}\ \bibnamefont {Niedzielski}}, \bibinfo {author}
  {\bibfnamefont {A.}~\bibnamefont {Karamlou}}, \bibinfo {author}
  {\bibfnamefont {K.}~\bibnamefont {Serniak}}, \bibinfo {author} {\bibfnamefont
  {A.}~\bibnamefont {Veps\"al\"ainen}}, \bibinfo {author} {\bibfnamefont
  {M.~E.}\ \bibnamefont {Schwartz}}, \bibinfo {author} {\bibfnamefont {J.~L.}\
  \bibnamefont {Yoder}}, \bibinfo {author} {\bibfnamefont {R.}~\bibnamefont
  {Winik}}, \bibinfo {author} {\bibfnamefont {J.~I.-J.}\ \bibnamefont {Wang}},
  \bibinfo {author} {\bibfnamefont {T.~P.}\ \bibnamefont {Orlando}}, \bibinfo
  {author} {\bibfnamefont {S.}~\bibnamefont {Gustavsson}}, \bibinfo {author}
  {\bibfnamefont {J.~A.}\ \bibnamefont {Grover}},\ and\ \bibinfo {author}
  {\bibfnamefont {W.~D.}\ \bibnamefont {Oliver}},\ }\bibfield  {title}
  {\bibinfo {title} {On-demand directional photon emission using waveguide
  quantum electrodynamics},\ }\href {https://doi.org/10.48550/arXiv.2203.01430}
  {\bibfield  {journal} {\bibinfo  {journal} {arXiv:2203.01430}\ } (\bibinfo
  {year} {2022})}\BibitemShut {NoStop}%
\bibitem [{\citenamefont {Poshakinskiy}\ and\ \citenamefont
  {Poddubny}(2019)}]{poshakinskiy2019optomechanical}%
  \BibitemOpen
  \bibfield  {author} {\bibinfo {author} {\bibfnamefont {A.}~\bibnamefont
  {Poshakinskiy}}\ and\ \bibinfo {author} {\bibfnamefont {A.}~\bibnamefont
  {Poddubny}},\ }\bibfield  {title} {\bibinfo {title} {Optomechanical {K}erker
  effect},\ }\href {https://doi.org/10.1103/PhysRevX.9.011008} {\bibfield
  {journal} {\bibinfo  {journal} {Physical Review X}\ }\textbf {\bibinfo
  {volume} {9}},\ \bibinfo {pages} {011008} (\bibinfo {year}
  {2019})}\BibitemShut {NoStop}%
\bibitem [{\citenamefont {Mirhosseini}\ \emph {et~al.}(2019)\citenamefont
  {Mirhosseini}, \citenamefont {Kim}, \citenamefont {Zhang}, \citenamefont
  {Sipahigil}, \citenamefont {Dieterle}, \citenamefont {Keller}, \citenamefont
  {Asenjo-Garcia}, \citenamefont {Chang},\ and\ \citenamefont
  {Painter}}]{mirhosseini2019cavity}%
  \BibitemOpen
  \bibfield  {author} {\bibinfo {author} {\bibfnamefont {M.}~\bibnamefont
  {Mirhosseini}}, \bibinfo {author} {\bibfnamefont {E.}~\bibnamefont {Kim}},
  \bibinfo {author} {\bibfnamefont {X.}~\bibnamefont {Zhang}}, \bibinfo
  {author} {\bibfnamefont {A.}~\bibnamefont {Sipahigil}}, \bibinfo {author}
  {\bibfnamefont {P.~B.}\ \bibnamefont {Dieterle}}, \bibinfo {author}
  {\bibfnamefont {A.~J.}\ \bibnamefont {Keller}}, \bibinfo {author}
  {\bibfnamefont {A.}~\bibnamefont {Asenjo-Garcia}}, \bibinfo {author}
  {\bibfnamefont {D.~E.}\ \bibnamefont {Chang}},\ and\ \bibinfo {author}
  {\bibfnamefont {O.}~\bibnamefont {Painter}},\ }\bibfield  {title} {\bibinfo
  {title} {Cavity quantum electrodynamics with atom-like mirrors},\ }\href
  {https://doi.org/10.1038/s41586-019-1196-1} {\bibfield  {journal} {\bibinfo
  {journal} {Nature}\ }\textbf {\bibinfo {volume} {569}},\ \bibinfo {pages}
  {692} (\bibinfo {year} {2019})}\BibitemShut {NoStop}%
\bibitem [{\citenamefont {Ozawa}\ \emph {et~al.}(2019)\citenamefont {Ozawa},
  \citenamefont {Price}, \citenamefont {Amo}, \citenamefont {Goldman},
  \citenamefont {Hafezi}, \citenamefont {Lu}, \citenamefont {Rechtsman},
  \citenamefont {Schuster}, \citenamefont {Simon}, \citenamefont {Zilberberg},\
  and\ \citenamefont {Carusotto}}]{Tomoki2019}%
  \BibitemOpen
  \bibfield  {author} {\bibinfo {author} {\bibfnamefont {T.}~\bibnamefont
  {Ozawa}}, \bibinfo {author} {\bibfnamefont {H.~M.}\ \bibnamefont {Price}},
  \bibinfo {author} {\bibfnamefont {A.}~\bibnamefont {Amo}}, \bibinfo {author}
  {\bibfnamefont {N.}~\bibnamefont {Goldman}}, \bibinfo {author} {\bibfnamefont
  {M.}~\bibnamefont {Hafezi}}, \bibinfo {author} {\bibfnamefont
  {L.}~\bibnamefont {Lu}}, \bibinfo {author} {\bibfnamefont {M.~C.}\
  \bibnamefont {Rechtsman}}, \bibinfo {author} {\bibfnamefont {D.}~\bibnamefont
  {Schuster}}, \bibinfo {author} {\bibfnamefont {J.}~\bibnamefont {Simon}},
  \bibinfo {author} {\bibfnamefont {O.}~\bibnamefont {Zilberberg}},\ and\
  \bibinfo {author} {\bibfnamefont {I.}~\bibnamefont {Carusotto}},\ }\bibfield
  {title} {\bibinfo {title} {Topological photonics},\ }\href
  {https://doi.org/10.1103/RevModPhys.91.015006} {\bibfield  {journal}
  {\bibinfo  {journal} {Rev. Mod. Phys.}\ }\textbf {\bibinfo {volume} {91}},\
  \bibinfo {pages} {015006} (\bibinfo {year} {2019})}\BibitemShut {NoStop}%
\bibitem [{\citenamefont {Kerker}\ \emph {et~al.}(1983)\citenamefont {Kerker},
  \citenamefont {Wang},\ and\ \citenamefont
  {Giles}}]{kerker1983electromagnetic}%
  \BibitemOpen
  \bibfield  {author} {\bibinfo {author} {\bibfnamefont {M.}~\bibnamefont
  {Kerker}}, \bibinfo {author} {\bibfnamefont {D.-S.}\ \bibnamefont {Wang}},\
  and\ \bibinfo {author} {\bibfnamefont {C.}~\bibnamefont {Giles}},\ }\bibfield
   {title} {\bibinfo {title} {Electromagnetic scattering by magnetic spheres},\
  }\href {https://doi.org/10.1364/JOSA.73.000765} {\bibfield  {journal}
  {\bibinfo  {journal} {JOSA}\ }\textbf {\bibinfo {volume} {73}},\ \bibinfo
  {pages} {765} (\bibinfo {year} {1983})}\BibitemShut {NoStop}%
\bibitem [{\citenamefont {Liu}\ and\ \citenamefont {Kivshar}(2018)}]{Liu2018}%
  \BibitemOpen
  \bibfield  {author} {\bibinfo {author} {\bibfnamefont {W.}~\bibnamefont
  {Liu}}\ and\ \bibinfo {author} {\bibfnamefont {Y.~S.}\ \bibnamefont
  {Kivshar}},\ }\bibfield  {title} {\bibinfo {title} {Generalized {K}erker
  effects in nanophotonics and meta-optics [invited]},\ }\href
  {https://doi.org/10.1364/oe.26.013085} {\bibfield  {journal} {\bibinfo
  {journal} {Optics Express}\ }\textbf {\bibinfo {volume} {26}},\ \bibinfo
  {pages} {13085} (\bibinfo {year} {2018})}\BibitemShut {NoStop}%
\bibitem [{\citenamefont {Staude}\ \emph {et~al.}(2013)\citenamefont {Staude},
  \citenamefont {Miroshnichenko}, \citenamefont {Decker}, \citenamefont
  {Fofang}, \citenamefont {Liu}, \citenamefont {Gonzales}, \citenamefont
  {Dominguez}, \citenamefont {Luk}, \citenamefont {Neshev}, \citenamefont
  {Brener},\ and\ \citenamefont {Kivshar}}]{Staude2013}%
  \BibitemOpen
  \bibfield  {author} {\bibinfo {author} {\bibfnamefont {I.}~\bibnamefont
  {Staude}}, \bibinfo {author} {\bibfnamefont {A.~E.}\ \bibnamefont
  {Miroshnichenko}}, \bibinfo {author} {\bibfnamefont {M.}~\bibnamefont
  {Decker}}, \bibinfo {author} {\bibfnamefont {N.~T.}\ \bibnamefont {Fofang}},
  \bibinfo {author} {\bibfnamefont {S.}~\bibnamefont {Liu}}, \bibinfo {author}
  {\bibfnamefont {E.}~\bibnamefont {Gonzales}}, \bibinfo {author}
  {\bibfnamefont {J.}~\bibnamefont {Dominguez}}, \bibinfo {author}
  {\bibfnamefont {T.~S.}\ \bibnamefont {Luk}}, \bibinfo {author} {\bibfnamefont
  {D.~N.}\ \bibnamefont {Neshev}}, \bibinfo {author} {\bibfnamefont
  {I.}~\bibnamefont {Brener}},\ and\ \bibinfo {author} {\bibfnamefont
  {Y.}~\bibnamefont {Kivshar}},\ }\bibfield  {title} {\bibinfo {title}
  {{T}ailoring {D}irectional {S}cattering through {M}agnetic and {E}lectric
  {R}esonances in {S}ubwavelength {S}ilicon {N}anodisks},\ }\href
  {http://pubs.acs.org/doi/abs/10.1021/nn402736f} {\bibfield  {journal}
  {\bibinfo  {journal} {ACS Nano}\ }\textbf {\bibinfo {volume} {7}},\ \bibinfo
  {pages} {7824} (\bibinfo {year} {2013})}\BibitemShut {NoStop}%
\bibitem [{\citenamefont {Fu}\ \emph {et~al.}(2013)\citenamefont {Fu},
  \citenamefont {Kuznetsov}, \citenamefont {Miroshnichenko}, \citenamefont
  {Yu},\ and\ \citenamefont {Luk'yanchuk}}]{fu2013directional}%
  \BibitemOpen
  \bibfield  {author} {\bibinfo {author} {\bibfnamefont {Y.~H.}\ \bibnamefont
  {Fu}}, \bibinfo {author} {\bibfnamefont {A.~I.}\ \bibnamefont {Kuznetsov}},
  \bibinfo {author} {\bibfnamefont {A.~E.}\ \bibnamefont {Miroshnichenko}},
  \bibinfo {author} {\bibfnamefont {Y.~F.}\ \bibnamefont {Yu}},\ and\ \bibinfo
  {author} {\bibfnamefont {B.}~\bibnamefont {Luk'yanchuk}},\ }\bibfield
  {title} {\bibinfo {title} {Directional visible light scattering by silicon
  nanoparticles},\ }\href {https://doi.org/10.1038/ncomms2538} {\bibfield
  {journal} {\bibinfo  {journal} {Nature communications}\ }\textbf {\bibinfo
  {volume} {4}},\ \bibinfo {pages} {1} (\bibinfo {year} {2013})}\BibitemShut
  {NoStop}%
\bibitem [{\citenamefont {Person}\ \emph {et~al.}(2013)\citenamefont {Person},
  \citenamefont {Jain}, \citenamefont {Lapin}, \citenamefont {S{\'a}enz},
  \citenamefont {Wicks},\ and\ \citenamefont
  {Novotny}}]{person2013demonstration}%
  \BibitemOpen
  \bibfield  {author} {\bibinfo {author} {\bibfnamefont {S.}~\bibnamefont
  {Person}}, \bibinfo {author} {\bibfnamefont {M.}~\bibnamefont {Jain}},
  \bibinfo {author} {\bibfnamefont {Z.}~\bibnamefont {Lapin}}, \bibinfo
  {author} {\bibfnamefont {J.~J.}\ \bibnamefont {S{\'a}enz}}, \bibinfo {author}
  {\bibfnamefont {G.}~\bibnamefont {Wicks}},\ and\ \bibinfo {author}
  {\bibfnamefont {L.}~\bibnamefont {Novotny}},\ }\bibfield  {title} {\bibinfo
  {title} {Demonstration of zero optical backscattering from single
  nanoparticles},\ }\href {https://doi.org/10.1021/nl4005018} {\bibfield
  {journal} {\bibinfo  {journal} {Nano Letters}\ }\textbf {\bibinfo {volume}
  {13}},\ \bibinfo {pages} {1806} (\bibinfo {year} {2013})}\BibitemShut
  {NoStop}%
\bibitem [{\citenamefont {Scheucher}\ \emph {et~al.}(2016)\citenamefont
  {Scheucher}, \citenamefont {Hilico}, \citenamefont {Will}, \citenamefont
  {Volz},\ and\ \citenamefont {Rauschenbeutel}}]{scheucher2016quantum}%
  \BibitemOpen
  \bibfield  {author} {\bibinfo {author} {\bibfnamefont {M.}~\bibnamefont
  {Scheucher}}, \bibinfo {author} {\bibfnamefont {A.}~\bibnamefont {Hilico}},
  \bibinfo {author} {\bibfnamefont {E.}~\bibnamefont {Will}}, \bibinfo {author}
  {\bibfnamefont {J.}~\bibnamefont {Volz}},\ and\ \bibinfo {author}
  {\bibfnamefont {A.}~\bibnamefont {Rauschenbeutel}},\ }\bibfield  {title}
  {\bibinfo {title} {Quantum optical circulator controlled by a single chirally
  coupled atom},\ }\href {DOI: 10.1126/science.aaj2118} {\bibfield  {journal}
  {\bibinfo  {journal} {Science}\ }\textbf {\bibinfo {volume} {354}},\ \bibinfo
  {pages} {1577} (\bibinfo {year} {2016})}\BibitemShut {NoStop}%
\bibitem [{\citenamefont {Lodahl}\ \emph {et~al.}(2017)\citenamefont {Lodahl},
  \citenamefont {Mahmoodian}, \citenamefont {Stobbe}, \citenamefont
  {Rauschenbeutel}, \citenamefont {Schneeweiss}, \citenamefont {Volz},
  \citenamefont {Pichler},\ and\ \citenamefont {Zoller}}]{lodahl2017chiral}%
  \BibitemOpen
  \bibfield  {author} {\bibinfo {author} {\bibfnamefont {P.}~\bibnamefont
  {Lodahl}}, \bibinfo {author} {\bibfnamefont {S.}~\bibnamefont {Mahmoodian}},
  \bibinfo {author} {\bibfnamefont {S.}~\bibnamefont {Stobbe}}, \bibinfo
  {author} {\bibfnamefont {A.}~\bibnamefont {Rauschenbeutel}}, \bibinfo
  {author} {\bibfnamefont {P.}~\bibnamefont {Schneeweiss}}, \bibinfo {author}
  {\bibfnamefont {J.}~\bibnamefont {Volz}}, \bibinfo {author} {\bibfnamefont
  {H.}~\bibnamefont {Pichler}},\ and\ \bibinfo {author} {\bibfnamefont
  {P.}~\bibnamefont {Zoller}},\ }\bibfield  {title} {\bibinfo {title} {Chiral
  quantum optics},\ }\href {https://doi.org/10.1038/nature21037} {\bibfield
  {journal} {\bibinfo  {journal} {Nature}\ }\textbf {\bibinfo {volume} {541}},\
  \bibinfo {pages} {473} (\bibinfo {year} {2017})}\BibitemShut {NoStop}%
\bibitem [{\citenamefont {Prasad}\ \emph {et~al.}(2020)\citenamefont {Prasad},
  \citenamefont {Hinney}, \citenamefont {Mahmoodian}, \citenamefont {Hammerer},
  \citenamefont {Rind}, \citenamefont {Schneeweiss}, \citenamefont
  {S{\o}rensen}, \citenamefont {Volz},\ and\ \citenamefont
  {Rauschenbeutel}}]{Prasad2020}%
  \BibitemOpen
  \bibfield  {author} {\bibinfo {author} {\bibfnamefont {A.~S.}\ \bibnamefont
  {Prasad}}, \bibinfo {author} {\bibfnamefont {J.}~\bibnamefont {Hinney}},
  \bibinfo {author} {\bibfnamefont {S.}~\bibnamefont {Mahmoodian}}, \bibinfo
  {author} {\bibfnamefont {K.}~\bibnamefont {Hammerer}}, \bibinfo {author}
  {\bibfnamefont {S.}~\bibnamefont {Rind}}, \bibinfo {author} {\bibfnamefont
  {P.}~\bibnamefont {Schneeweiss}}, \bibinfo {author} {\bibfnamefont {A.~S.}\
  \bibnamefont {S{\o}rensen}}, \bibinfo {author} {\bibfnamefont
  {J.}~\bibnamefont {Volz}},\ and\ \bibinfo {author} {\bibfnamefont
  {A.}~\bibnamefont {Rauschenbeutel}},\ }\bibfield  {title} {\bibinfo {title}
  {Correlating photons using the collective nonlinear response of atoms weakly
  coupled to an optical mode},\ }\href
  {https://doi.org/10.1038/s41566-020-0692-z} {\bibfield  {journal} {\bibinfo
  {journal} {Nature Photonics}\ }\textbf {\bibinfo {volume} {14}},\ \bibinfo
  {pages} {719} (\bibinfo {year} {2020})}\BibitemShut {NoStop}%
\bibitem [{\citenamefont {Zhang}\ \emph {et~al.}(2018)\citenamefont {Zhang},
  \citenamefont {Hu}, \citenamefont {Lin}, \citenamefont {Niu}, \citenamefont
  {Xia}, \citenamefont {Gong},\ and\ \citenamefont {Gong}}]{zhang2018thermal}%
  \BibitemOpen
  \bibfield  {author} {\bibinfo {author} {\bibfnamefont {S.}~\bibnamefont
  {Zhang}}, \bibinfo {author} {\bibfnamefont {Y.}~\bibnamefont {Hu}}, \bibinfo
  {author} {\bibfnamefont {G.}~\bibnamefont {Lin}}, \bibinfo {author}
  {\bibfnamefont {Y.}~\bibnamefont {Niu}}, \bibinfo {author} {\bibfnamefont
  {K.}~\bibnamefont {Xia}}, \bibinfo {author} {\bibfnamefont {J.}~\bibnamefont
  {Gong}},\ and\ \bibinfo {author} {\bibfnamefont {S.}~\bibnamefont {Gong}},\
  }\bibfield  {title} {\bibinfo {title} {Thermal-motion-induced non-reciprocal
  quantum optical system},\ }\href {https://doi.org/10.1038/s41566-018-0269-2}
  {\bibfield  {journal} {\bibinfo  {journal} {Nature Photonics}\ }\textbf
  {\bibinfo {volume} {12}},\ \bibinfo {pages} {744} (\bibinfo {year}
  {2018})}\BibitemShut {NoStop}%
\bibitem [{\citenamefont {M{\"u}ller}\ \emph {et~al.}(2010)\citenamefont
  {M{\"u}ller}, \citenamefont {Krause}, \citenamefont {Renner},\ and\
  \citenamefont {Brinkmeyer}}]{muller2010measurement}%
  \BibitemOpen
  \bibfield  {author} {\bibinfo {author} {\bibfnamefont {J.}~\bibnamefont
  {M{\"u}ller}}, \bibinfo {author} {\bibfnamefont {M.}~\bibnamefont {Krause}},
  \bibinfo {author} {\bibfnamefont {H.}~\bibnamefont {Renner}},\ and\ \bibinfo
  {author} {\bibfnamefont {E.}~\bibnamefont {Brinkmeyer}},\ }\bibfield  {title}
  {\bibinfo {title} {Measurement of nonreciprocal spontaneous {R}aman
  scattering in silicon photonic wires},\ }\href
  {https://doi.org/10.1364/OE.18.019532} {\bibfield  {journal} {\bibinfo
  {journal} {Optics Express}\ }\textbf {\bibinfo {volume} {18}},\ \bibinfo
  {pages} {19532} (\bibinfo {year} {2010})}\BibitemShut {NoStop}%
\bibitem [{\citenamefont {Kittlaus}\ \emph {et~al.}(2018)\citenamefont
  {Kittlaus}, \citenamefont {Otterstrom}, \citenamefont {Kharel}, \citenamefont
  {Gertler},\ and\ \citenamefont {Rakich}}]{kittlaus2018non}%
  \BibitemOpen
  \bibfield  {author} {\bibinfo {author} {\bibfnamefont {E.~A.}\ \bibnamefont
  {Kittlaus}}, \bibinfo {author} {\bibfnamefont {N.~T.}\ \bibnamefont
  {Otterstrom}}, \bibinfo {author} {\bibfnamefont {P.}~\bibnamefont {Kharel}},
  \bibinfo {author} {\bibfnamefont {S.}~\bibnamefont {Gertler}},\ and\ \bibinfo
  {author} {\bibfnamefont {P.~T.}\ \bibnamefont {Rakich}},\ }\bibfield  {title}
  {\bibinfo {title} {Non-reciprocal interband {B}rillouin modulation},\ }\href
  {https://doi.org/10.1038/s41566-018-0254-9} {\bibfield  {journal} {\bibinfo
  {journal} {Nature Photonics}\ }\textbf {\bibinfo {volume} {12}},\ \bibinfo
  {pages} {613} (\bibinfo {year} {2018})}\BibitemShut {NoStop}%
\bibitem [{\citenamefont {Guimond}\ \emph {et~al.}(2020)\citenamefont
  {Guimond}, \citenamefont {Vermersch}, \citenamefont {Juan}, \citenamefont
  {Sharafiev}, \citenamefont {Kirchmair},\ and\ \citenamefont
  {Zoller}}]{Guimond2020}%
  \BibitemOpen
  \bibfield  {author} {\bibinfo {author} {\bibfnamefont {P.-O.}\ \bibnamefont
  {Guimond}}, \bibinfo {author} {\bibfnamefont {B.}~\bibnamefont {Vermersch}},
  \bibinfo {author} {\bibfnamefont {M.~L.}\ \bibnamefont {Juan}}, \bibinfo
  {author} {\bibfnamefont {A.}~\bibnamefont {Sharafiev}}, \bibinfo {author}
  {\bibfnamefont {G.}~\bibnamefont {Kirchmair}},\ and\ \bibinfo {author}
  {\bibfnamefont {P.}~\bibnamefont {Zoller}},\ }\bibfield  {title} {\bibinfo
  {title} {A unidirectional on-chip photonic interface for superconducting
  circuits},\ }\href {https://doi.org/10.1038/s41534-020-0261-9} {\bibfield
  {journal} {\bibinfo  {journal} {npj Quantum Information}\ }\textbf {\bibinfo
  {volume} {6}},\ \bibinfo {pages} {32} (\bibinfo {year} {2020})}\BibitemShut
  {NoStop}%
\bibitem [{\citenamefont {Mahoney}\ \emph {et~al.}(2017)\citenamefont
  {Mahoney}, \citenamefont {Colless}, \citenamefont {Pauka}, \citenamefont
  {Hornibrook}, \citenamefont {Watson}, \citenamefont {Gardner}, \citenamefont
  {Manfra}, \citenamefont {Doherty},\ and\ \citenamefont
  {Reilly}}]{mahoney2017chip}%
  \BibitemOpen
  \bibfield  {author} {\bibinfo {author} {\bibfnamefont {A.}~\bibnamefont
  {Mahoney}}, \bibinfo {author} {\bibfnamefont {J.}~\bibnamefont {Colless}},
  \bibinfo {author} {\bibfnamefont {S.}~\bibnamefont {Pauka}}, \bibinfo
  {author} {\bibfnamefont {J.}~\bibnamefont {Hornibrook}}, \bibinfo {author}
  {\bibfnamefont {J.}~\bibnamefont {Watson}}, \bibinfo {author} {\bibfnamefont
  {G.}~\bibnamefont {Gardner}}, \bibinfo {author} {\bibfnamefont
  {M.}~\bibnamefont {Manfra}}, \bibinfo {author} {\bibfnamefont
  {A.}~\bibnamefont {Doherty}},\ and\ \bibinfo {author} {\bibfnamefont
  {D.}~\bibnamefont {Reilly}},\ }\bibfield  {title} {\bibinfo {title} {On-chip
  microwave quantum {H}all circulator},\ }\href
  {https://doi.org/10.1103/PhysRevX.7.011007} {\bibfield  {journal} {\bibinfo
  {journal} {Physical Review X}\ }\textbf {\bibinfo {volume} {7}},\ \bibinfo
  {pages} {011007} (\bibinfo {year} {2017})}\BibitemShut {NoStop}%
\bibitem [{\citenamefont {Peterson}\ \emph {et~al.}(2017)\citenamefont
  {Peterson}, \citenamefont {Lecocq}, \citenamefont {Cicak}, \citenamefont
  {Simmonds}, \citenamefont {Aumentado},\ and\ \citenamefont
  {Teufel}}]{peterson2017demonstration}%
  \BibitemOpen
  \bibfield  {author} {\bibinfo {author} {\bibfnamefont {G.~A.}\ \bibnamefont
  {Peterson}}, \bibinfo {author} {\bibfnamefont {F.}~\bibnamefont {Lecocq}},
  \bibinfo {author} {\bibfnamefont {K.}~\bibnamefont {Cicak}}, \bibinfo
  {author} {\bibfnamefont {R.~W.}\ \bibnamefont {Simmonds}}, \bibinfo {author}
  {\bibfnamefont {J.}~\bibnamefont {Aumentado}},\ and\ \bibinfo {author}
  {\bibfnamefont {J.~D.}\ \bibnamefont {Teufel}},\ }\bibfield  {title}
  {\bibinfo {title} {Demonstration of efficient nonreciprocity in a microwave
  optomechanical circuit},\ }\href {https://doi.org/10.1103/PhysRevX.7.031001}
  {\bibfield  {journal} {\bibinfo  {journal} {Physical Review X}\ }\textbf
  {\bibinfo {volume} {7}},\ \bibinfo {pages} {031001} (\bibinfo {year}
  {2017})}\BibitemShut {NoStop}%
\bibitem [{\citenamefont {Bernier}\ \emph {et~al.}(2017)\citenamefont
  {Bernier}, \citenamefont {T\'oth}, \citenamefont {Koottandavida},
  \citenamefont {Ioannou}, \citenamefont {Malz}, \citenamefont {Nunnenkamp},
  \citenamefont {Feofanov},\ and\ \citenamefont {Kippenberg}}]{bernier2017}%
  \BibitemOpen
  \bibfield  {author} {\bibinfo {author} {\bibfnamefont {N.~R.}\ \bibnamefont
  {Bernier}}, \bibinfo {author} {\bibfnamefont {L.~D.}\ \bibnamefont {T\'oth}},
  \bibinfo {author} {\bibfnamefont {A.}~\bibnamefont {Koottandavida}}, \bibinfo
  {author} {\bibfnamefont {M.~A.}\ \bibnamefont {Ioannou}}, \bibinfo {author}
  {\bibfnamefont {D.}~\bibnamefont {Malz}}, \bibinfo {author} {\bibfnamefont
  {A.}~\bibnamefont {Nunnenkamp}}, \bibinfo {author} {\bibfnamefont {A.~K.}\
  \bibnamefont {Feofanov}},\ and\ \bibinfo {author} {\bibfnamefont {T.~J.}\
  \bibnamefont {Kippenberg}},\ }\bibfield  {title} {\bibinfo {title}
  {Nonreciprocal reconfigurable microwave optomechanical circuit},\ }\href
  {https://www.nature.com/articles/s41467-017-00447-1} {\bibfield  {journal}
  {\bibinfo  {journal} {Nature Communications}\ }\textbf {\bibinfo {volume}
  {8}} (\bibinfo {year} {2017})}\BibitemShut {NoStop}%
\bibitem [{\citenamefont {Barzanjeh}\ \emph {et~al.}(2017)\citenamefont
  {Barzanjeh}, \citenamefont {Wulf}, \citenamefont {Peruzzo}, \citenamefont
  {Kalaee}, \citenamefont {Dieterle}, \citenamefont {Painter},\ and\
  \citenamefont {Fink}}]{barzanjeh2017mechanical}%
  \BibitemOpen
  \bibfield  {author} {\bibinfo {author} {\bibfnamefont {S.}~\bibnamefont
  {Barzanjeh}}, \bibinfo {author} {\bibfnamefont {M.}~\bibnamefont {Wulf}},
  \bibinfo {author} {\bibfnamefont {M.}~\bibnamefont {Peruzzo}}, \bibinfo
  {author} {\bibfnamefont {M.}~\bibnamefont {Kalaee}}, \bibinfo {author}
  {\bibfnamefont {P.}~\bibnamefont {Dieterle}}, \bibinfo {author}
  {\bibfnamefont {O.}~\bibnamefont {Painter}},\ and\ \bibinfo {author}
  {\bibfnamefont {J.~M.}\ \bibnamefont {Fink}},\ }\bibfield  {title} {\bibinfo
  {title} {Mechanical on-chip microwave circulator},\ }\href@noop {} {\bibfield
   {journal} {\bibinfo  {journal} {Nature communications}\ }\textbf {\bibinfo
  {volume} {8}},\ \bibinfo {pages} {1} (\bibinfo {year} {2017})}\BibitemShut
  {NoStop}%
\bibitem [{\citenamefont {Chapman}\ \emph {et~al.}(2017)\citenamefont
  {Chapman}, \citenamefont {Rosenthal}, \citenamefont {Kerckhoff},
  \citenamefont {Moores}, \citenamefont {Vale}, \citenamefont {Mates},
  \citenamefont {Hilton}, \citenamefont {Lalumiere}, \citenamefont {Blais},\
  and\ \citenamefont {Lehnert}}]{chapman2017widely}%
  \BibitemOpen
  \bibfield  {author} {\bibinfo {author} {\bibfnamefont {B.~J.}\ \bibnamefont
  {Chapman}}, \bibinfo {author} {\bibfnamefont {E.~I.}\ \bibnamefont
  {Rosenthal}}, \bibinfo {author} {\bibfnamefont {J.}~\bibnamefont
  {Kerckhoff}}, \bibinfo {author} {\bibfnamefont {B.~A.}\ \bibnamefont
  {Moores}}, \bibinfo {author} {\bibfnamefont {L.~R.}\ \bibnamefont {Vale}},
  \bibinfo {author} {\bibfnamefont {J.}~\bibnamefont {Mates}}, \bibinfo
  {author} {\bibfnamefont {G.~C.}\ \bibnamefont {Hilton}}, \bibinfo {author}
  {\bibfnamefont {K.}~\bibnamefont {Lalumiere}}, \bibinfo {author}
  {\bibfnamefont {A.}~\bibnamefont {Blais}},\ and\ \bibinfo {author}
  {\bibfnamefont {K.}~\bibnamefont {Lehnert}},\ }\bibfield  {title} {\bibinfo
  {title} {Widely tunable on-chip microwave circulator for superconducting
  quantum circuits},\ }\href {https://doi.org/10.1103/PhysRevX.7.041043}
  {\bibfield  {journal} {\bibinfo  {journal} {Physical Review X}\ }\textbf
  {\bibinfo {volume} {7}},\ \bibinfo {pages} {041043} (\bibinfo {year}
  {2017})}\BibitemShut {NoStop}%
\bibitem [{\citenamefont {Rosenthal}\ \emph {et~al.}(2017)\citenamefont
  {Rosenthal}, \citenamefont {Chapman}, \citenamefont {Higginbotham},
  \citenamefont {Kerckhoff},\ and\ \citenamefont
  {Lehnert}}]{rosenthal2017breaking}%
  \BibitemOpen
  \bibfield  {author} {\bibinfo {author} {\bibfnamefont {E.~I.}\ \bibnamefont
  {Rosenthal}}, \bibinfo {author} {\bibfnamefont {B.~J.}\ \bibnamefont
  {Chapman}}, \bibinfo {author} {\bibfnamefont {A.~P.}\ \bibnamefont
  {Higginbotham}}, \bibinfo {author} {\bibfnamefont {J.}~\bibnamefont
  {Kerckhoff}},\ and\ \bibinfo {author} {\bibfnamefont {K.~W.}\ \bibnamefont
  {Lehnert}},\ }\bibfield  {title} {\bibinfo {title} {Breaking {L}orentz
  reciprocity with frequency conversion and delay},\ }\href
  {https://doi.org/10.1103/PhysRevLett.119.147703} {\bibfield  {journal}
  {\bibinfo  {journal} {Phys. Rev. Lett.}\ }\textbf {\bibinfo {volume} {119}},\
  \bibinfo {pages} {147703} (\bibinfo {year} {2017})}\BibitemShut {NoStop}%
\bibitem [{\citenamefont {Liu}\ \emph {et~al.}(2018)\citenamefont {Liu},
  \citenamefont {Powell}, \citenamefont {Zarate},\ and\ \citenamefont
  {Shadrivov}}]{Shadrivov2018}%
  \BibitemOpen
  \bibfield  {author} {\bibinfo {author} {\bibfnamefont {M.}~\bibnamefont
  {Liu}}, \bibinfo {author} {\bibfnamefont {D.~A.}\ \bibnamefont {Powell}},
  \bibinfo {author} {\bibfnamefont {Y.}~\bibnamefont {Zarate}},\ and\ \bibinfo
  {author} {\bibfnamefont {I.~V.}\ \bibnamefont {Shadrivov}},\ }\bibfield
  {title} {\bibinfo {title} {Huygens' metadevices for parametric waves},\
  }\href {https://doi.org/10.1103/PhysRevX.8.031077} {\bibfield  {journal}
  {\bibinfo  {journal} {Phys. Rev. X}\ }\textbf {\bibinfo {volume} {8}},\
  \bibinfo {pages} {031077} (\bibinfo {year} {2018})}\BibitemShut {NoStop}%
\bibitem [{\citenamefont {Jackel}\ and\ \citenamefont
  {Reynaud}(1992)}]{jackel1992fluctuations}%
  \BibitemOpen
  \bibfield  {author} {\bibinfo {author} {\bibfnamefont {M.-T.}\ \bibnamefont
  {Jackel}}\ and\ \bibinfo {author} {\bibfnamefont {S.}~\bibnamefont
  {Reynaud}},\ }\bibfield  {title} {\bibinfo {title} {Fluctuations and
  dissipation for a mirror in vacuum},\ }\href
  {https://doi.org/10.1088/0954-8998/4/1/005} {\bibfield  {journal} {\bibinfo
  {journal} {Quantum Optics: Journal of the European Optical Society Part B}\
  }\textbf {\bibinfo {volume} {4}},\ \bibinfo {pages} {39} (\bibinfo {year}
  {1992})}\BibitemShut {NoStop}%
\bibitem [{\citenamefont {Kannan}\ \emph {et~al.}(2020)\citenamefont {Kannan},
  \citenamefont {Campbell}, \citenamefont {Vasconcelos}, \citenamefont {Winik},
  \citenamefont {Kim}, \citenamefont {Kjaergaard}, \citenamefont {Krantz},
  \citenamefont {Melville}, \citenamefont {Niedzielski}, \citenamefont {Yoder},
  \citenamefont {Orlando}, \citenamefont {Gustavsson},\ and\ \citenamefont
  {Oliver}}]{oliver2020b}%
  \BibitemOpen
  \bibfield  {author} {\bibinfo {author} {\bibfnamefont {B.}~\bibnamefont
  {Kannan}}, \bibinfo {author} {\bibfnamefont {D.~L.}\ \bibnamefont
  {Campbell}}, \bibinfo {author} {\bibfnamefont {F.}~\bibnamefont
  {Vasconcelos}}, \bibinfo {author} {\bibfnamefont {R.}~\bibnamefont {Winik}},
  \bibinfo {author} {\bibfnamefont {D.~K.}\ \bibnamefont {Kim}}, \bibinfo
  {author} {\bibfnamefont {M.}~\bibnamefont {Kjaergaard}}, \bibinfo {author}
  {\bibfnamefont {P.}~\bibnamefont {Krantz}}, \bibinfo {author} {\bibfnamefont
  {A.}~\bibnamefont {Melville}}, \bibinfo {author} {\bibfnamefont {B.~M.}\
  \bibnamefont {Niedzielski}}, \bibinfo {author} {\bibfnamefont {J.~L.}\
  \bibnamefont {Yoder}}, \bibinfo {author} {\bibfnamefont {T.~P.}\ \bibnamefont
  {Orlando}}, \bibinfo {author} {\bibfnamefont {S.}~\bibnamefont
  {Gustavsson}},\ and\ \bibinfo {author} {\bibfnamefont {W.~D.}\ \bibnamefont
  {Oliver}},\ }\bibfield  {title} {\bibinfo {title} {Generating spatially
  entangled itinerant photons with waveguide quantum electrodynamics},\ }\href
  {https://doi.org/10.1126/sciadv.abb8780} {\bibfield  {journal} {\bibinfo
  {journal} {Science Advances}\ }\textbf {\bibinfo {volume} {6}},\ \bibinfo
  {pages} {eabb8780} (\bibinfo {year} {2020})}\BibitemShut {NoStop}%
\bibitem [{\citenamefont {Gheeraert}\ \emph {et~al.}(2020)\citenamefont
  {Gheeraert}, \citenamefont {Kono},\ and\ \citenamefont
  {Nakamura}}]{Gheeraert2020}%
  \BibitemOpen
  \bibfield  {author} {\bibinfo {author} {\bibfnamefont {N.}~\bibnamefont
  {Gheeraert}}, \bibinfo {author} {\bibfnamefont {S.}~\bibnamefont {Kono}},\
  and\ \bibinfo {author} {\bibfnamefont {Y.}~\bibnamefont {Nakamura}},\
  }\bibfield  {title} {\bibinfo {title} {Programmable directional emitter and
  receiver of itinerant microwave photons in a waveguide},\ }\href
  {https://doi.org/10.1103/PhysRevA.102.053720} {\bibfield  {journal} {\bibinfo
   {journal} {Phys. Rev. A}\ }\textbf {\bibinfo {volume} {102}},\ \bibinfo
  {pages} {053720} (\bibinfo {year} {2020})}\BibitemShut {NoStop}%
\bibitem [{\citenamefont {Poshakinskiy}\ and\ \citenamefont
  {Tarasenko}(2020)}]{poshakinskiy2020spin}%
  \BibitemOpen
  \bibfield  {author} {\bibinfo {author} {\bibfnamefont {A.}~\bibnamefont
  {Poshakinskiy}}\ and\ \bibinfo {author} {\bibfnamefont {S.}~\bibnamefont
  {Tarasenko}},\ }\bibfield  {title} {\bibinfo {title} {Spin noise at electron
  paramagnetic resonance},\ }\href
  {https://doi.org/10.1103/PhysRevB.101.075403} {\bibfield  {journal} {\bibinfo
   {journal} {Physical Review B}\ }\textbf {\bibinfo {volume} {101}},\ \bibinfo
  {pages} {075403} (\bibinfo {year} {2020})}\BibitemShut {NoStop}%
\bibitem [{\citenamefont {Lu}\ \emph {et~al.}(2016)\citenamefont {Lu},
  \citenamefont {Joannopoulos},\ and\ \citenamefont
  {Solja{\v{c}}i{\'c}}}]{lu2016topological}%
  \BibitemOpen
  \bibfield  {author} {\bibinfo {author} {\bibfnamefont {L.}~\bibnamefont
  {Lu}}, \bibinfo {author} {\bibfnamefont {J.~D.}\ \bibnamefont
  {Joannopoulos}},\ and\ \bibinfo {author} {\bibfnamefont {M.}~\bibnamefont
  {Solja{\v{c}}i{\'c}}},\ }\bibfield  {title} {\bibinfo {title} {Topological
  states in photonic systems},\ }\href {https://doi.org/10.1038/nphys3796}
  {\bibfield  {journal} {\bibinfo  {journal} {Nature Physics}\ }\textbf
  {\bibinfo {volume} {12}},\ \bibinfo {pages} {626} (\bibinfo {year}
  {2016})}\BibitemShut {NoStop}%
\bibitem [{\citenamefont {Rymarz}\ \emph {et~al.}(2021)\citenamefont {Rymarz},
  \citenamefont {Bosco}, \citenamefont {Ciani},\ and\ \citenamefont
  {DiVincenzo}}]{rymarz2021hardware}%
  \BibitemOpen
  \bibfield  {author} {\bibinfo {author} {\bibfnamefont {M.}~\bibnamefont
  {Rymarz}}, \bibinfo {author} {\bibfnamefont {S.}~\bibnamefont {Bosco}},
  \bibinfo {author} {\bibfnamefont {A.}~\bibnamefont {Ciani}},\ and\ \bibinfo
  {author} {\bibfnamefont {D.~P.}\ \bibnamefont {DiVincenzo}},\ }\bibfield
  {title} {\bibinfo {title} {Hardware-encoding grid states in a nonreciprocal
  superconducting circuit},\ }\href
  {https://doi.org/10.1103/PhysRevX.11.011032} {\bibfield  {journal} {\bibinfo
  {journal} {Phys. Rev. X}\ }\textbf {\bibinfo {volume} {11}},\ \bibinfo
  {pages} {011032} (\bibinfo {year} {2021})}\BibitemShut {NoStop}%
\bibitem [{\citenamefont {Caneva}\ \emph {et~al.}(2015)\citenamefont {Caneva},
  \citenamefont {Manzoni}, \citenamefont {Shi}, \citenamefont {Douglas},
  \citenamefont {Cirac},\ and\ \citenamefont {Chang}}]{Caneva2015}%
  \BibitemOpen
  \bibfield  {author} {\bibinfo {author} {\bibfnamefont {T.}~\bibnamefont
  {Caneva}}, \bibinfo {author} {\bibfnamefont {M.~T.}\ \bibnamefont {Manzoni}},
  \bibinfo {author} {\bibfnamefont {T.}~\bibnamefont {Shi}}, \bibinfo {author}
  {\bibfnamefont {J.~S.}\ \bibnamefont {Douglas}}, \bibinfo {author}
  {\bibfnamefont {J.~I.}\ \bibnamefont {Cirac}},\ and\ \bibinfo {author}
  {\bibfnamefont {D.~E.}\ \bibnamefont {Chang}},\ }\bibfield  {title} {\bibinfo
  {title} {Quantum dynamics of propagating photons with strong interactions: a
  generalized input{\textendash}output formalism},\ }\href
  {https://doi.org/10.1088/1367-2630/17/11/113001} {\bibfield  {journal}
  {\bibinfo  {journal} {New Journal of Physics}\ }\textbf {\bibinfo {volume}
  {17}},\ \bibinfo {pages} {113001} (\bibinfo {year} {2015})}\BibitemShut
  {NoStop}%
\bibitem [{\citenamefont {Poddubny}\ and\ \citenamefont
  {Golub}(2021)}]{Poddubny2021Ratchet}%
  \BibitemOpen
  \bibfield  {author} {\bibinfo {author} {\bibfnamefont {A.~N.}\ \bibnamefont
  {Poddubny}}\ and\ \bibinfo {author} {\bibfnamefont {L.~E.}\ \bibnamefont
  {Golub}},\ }\bibfield  {title} {\bibinfo {title} {Ratchet effect in
  frequency-modulated waveguide-coupled emitter arrays},\ }\href
  {https://doi.org/10.1103/PhysRevB.104.205309} {\bibfield  {journal} {\bibinfo
   {journal} {Phys. Rev. B}\ }\textbf {\bibinfo {volume} {104}},\ \bibinfo
  {pages} {205309} (\bibinfo {year} {2021})}\BibitemShut {NoStop}%
\bibitem [{\citenamefont {Marquardt}\ \emph {et~al.}(2006)\citenamefont
  {Marquardt}, \citenamefont {Harris},\ and\ \citenamefont
  {Girvin}}]{Marquardt2006}%
  \BibitemOpen
  \bibfield  {author} {\bibinfo {author} {\bibfnamefont {F.}~\bibnamefont
  {Marquardt}}, \bibinfo {author} {\bibfnamefont {J.~G.~E.}\ \bibnamefont
  {Harris}},\ and\ \bibinfo {author} {\bibfnamefont {S.~M.}\ \bibnamefont
  {Girvin}},\ }\bibfield  {title} {\bibinfo {title} {Dynamical multistability
  induced by radiation pressure in high-finesse micromechanical optical
  cavities},\ }\href {https://doi.org/10.1103/PhysRevLett.96.103901} {\bibfield
   {journal} {\bibinfo  {journal} {Phys. Rev. Lett.}\ }\textbf {\bibinfo
  {volume} {96}},\ \bibinfo {pages} {103901} (\bibinfo {year}
  {2006})}\BibitemShut {NoStop}%
\bibitem [{\citenamefont {Celi}\ \emph {et~al.}(2014)\citenamefont {Celi},
  \citenamefont {Massignan}, \citenamefont {Ruseckas}, \citenamefont {Goldman},
  \citenamefont {Spielman}, \citenamefont {Juzeli\ifmmode~\bar{u}\else
  \={u}\fi{}nas},\ and\ \citenamefont {Lewenstein}}]{Lewenstein2014}%
  \BibitemOpen
  \bibfield  {author} {\bibinfo {author} {\bibfnamefont {A.}~\bibnamefont
  {Celi}}, \bibinfo {author} {\bibfnamefont {P.}~\bibnamefont {Massignan}},
  \bibinfo {author} {\bibfnamefont {J.}~\bibnamefont {Ruseckas}}, \bibinfo
  {author} {\bibfnamefont {N.}~\bibnamefont {Goldman}}, \bibinfo {author}
  {\bibfnamefont {I.~B.}\ \bibnamefont {Spielman}}, \bibinfo {author}
  {\bibfnamefont {G.}~\bibnamefont {Juzeli\ifmmode~\bar{u}\else
  \={u}\fi{}nas}},\ and\ \bibinfo {author} {\bibfnamefont {M.}~\bibnamefont
  {Lewenstein}},\ }\bibfield  {title} {\bibinfo {title} {Synthetic gauge fields
  in synthetic dimensions},\ }\href
  {https://doi.org/10.1103/PhysRevLett.112.043001} {\bibfield  {journal}
  {\bibinfo  {journal} {Phys. Rev. Lett.}\ }\textbf {\bibinfo {volume} {112}},\
  \bibinfo {pages} {043001} (\bibinfo {year} {2014})}\BibitemShut {NoStop}%
\bibitem [{\citenamefont {Yuan}\ \emph {et~al.}(2016)\citenamefont {Yuan},
  \citenamefont {Shi},\ and\ \citenamefont {Fan}}]{Yuan:16}%
  \BibitemOpen
  \bibfield  {author} {\bibinfo {author} {\bibfnamefont {L.}~\bibnamefont
  {Yuan}}, \bibinfo {author} {\bibfnamefont {Y.}~\bibnamefont {Shi}},\ and\
  \bibinfo {author} {\bibfnamefont {S.}~\bibnamefont {Fan}},\ }\bibfield
  {title} {\bibinfo {title} {Photonic gauge potential in a system with a
  synthetic frequency dimension},\ }\href
  {https://doi.org/10.1364/OL.41.000741} {\bibfield  {journal} {\bibinfo
  {journal} {Opt. Lett.}\ }\textbf {\bibinfo {volume} {41}},\ \bibinfo {pages}
  {741} (\bibinfo {year} {2016})}\BibitemShut {NoStop}%
\bibitem [{\citenamefont {Lalumi\`ere}\ \emph {et~al.}(2013)\citenamefont
  {Lalumi\`ere}, \citenamefont {Sanders}, \citenamefont {van Loo},
  \citenamefont {Fedorov}, \citenamefont {Wallraff},\ and\ \citenamefont
  {Blais}}]{Blais2013}%
  \BibitemOpen
  \bibfield  {author} {\bibinfo {author} {\bibfnamefont {K.}~\bibnamefont
  {Lalumi\`ere}}, \bibinfo {author} {\bibfnamefont {B.~C.}\ \bibnamefont
  {Sanders}}, \bibinfo {author} {\bibfnamefont {A.~F.}\ \bibnamefont {van
  Loo}}, \bibinfo {author} {\bibfnamefont {A.}~\bibnamefont {Fedorov}},
  \bibinfo {author} {\bibfnamefont {A.}~\bibnamefont {Wallraff}},\ and\
  \bibinfo {author} {\bibfnamefont {A.}~\bibnamefont {Blais}},\ }\bibfield
  {title} {\bibinfo {title} {Input-output theory for waveguide {QED} with an
  ensemble of inhomogeneous atoms},\ }\href
  {https://doi.org/10.1103/PhysRevA.88.043806} {\bibfield  {journal} {\bibinfo
  {journal} {Phys. Rev. A}\ }\textbf {\bibinfo {volume} {88}},\ \bibinfo
  {pages} {043806} (\bibinfo {year} {2013})}\BibitemShut {NoStop}%
\end{thebibliography}%
\onecolumngrid
\newpage
\normalsize
\appendix
\begin{center}
\textbf{\large{Supplementary Information for: Tunable directional photon scattering from a pair of superconducting qubits}}
\end{center}

\section*{Supplementary discussion}

\subsection*{Gyrator properties}

At the relative phase $\alpha/\pi = \pm 1$, the scattering coefficients of inelastic transmission in different directions are equal in absolute value and have different signs, $S_{12}=-S_{21}$, ($\pi$ phase-shift) which is indeed one of the signatures of a gyrator with the ideal S-matrix 
\begin{eqnarray}\label{SM_gy}
&& S_{\rm gyrator} = \begin{pmatrix}
0 & 1 \\
-1 & 0 
\end{pmatrix}.
\end{eqnarray}
For the presented device the nonreciprocal phase shift is robust for variable system parameters as it only depends on the correct relative phase between the two modulation tones. However, at $\alpha/\pi = \pm 1$, our device mostly scatters backward, which means that $|S_{12(21)}| \ll |S_{11(22)}|$ and the gyrator efficiency is rather low.

\subsection*{Isolator properties}
\begin{figure}[h]
\centering
\includegraphics[width=.35\columnwidth]{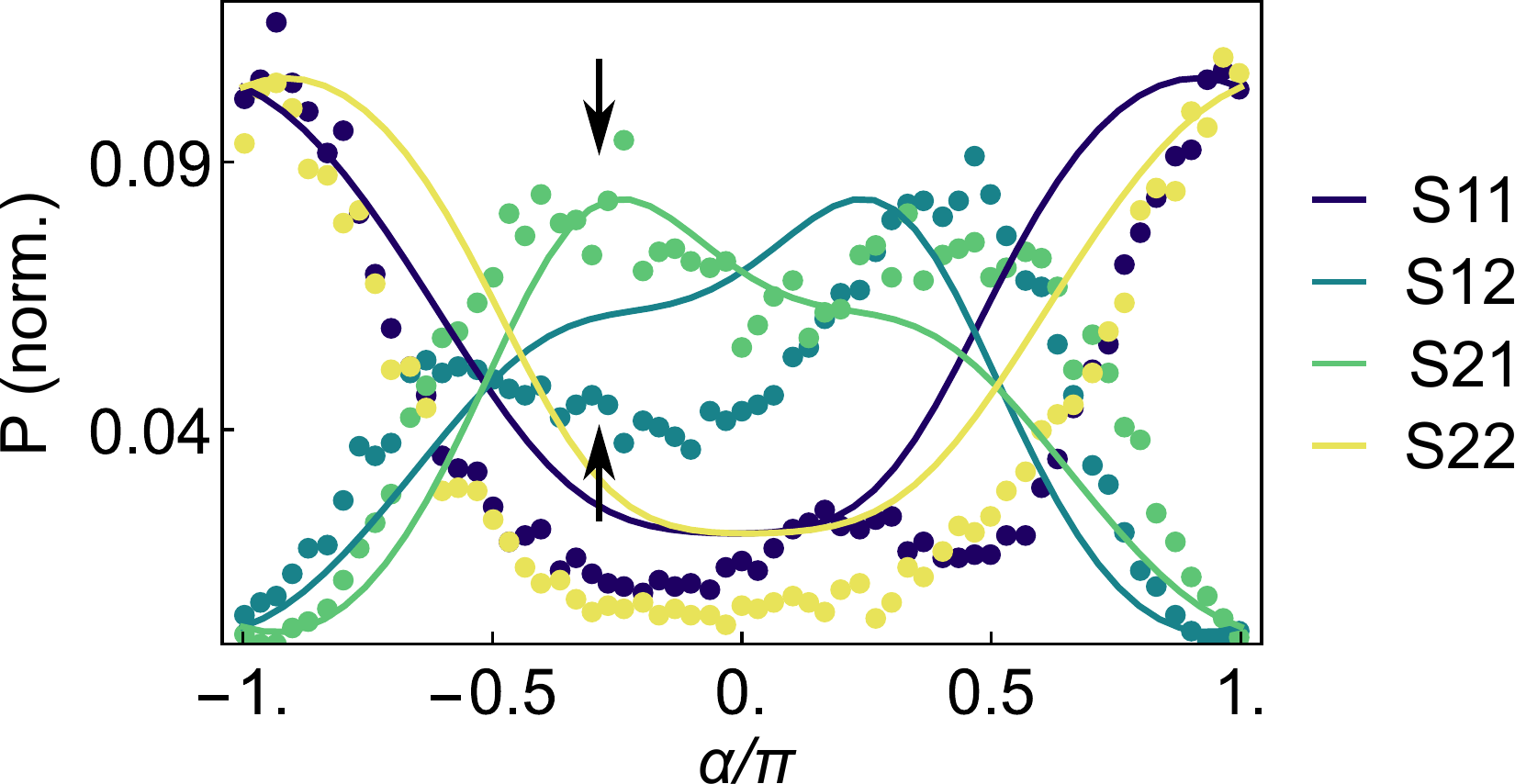}
\caption{\textbf{Isolator scattering matrix.} Measured coherent inelastic scattering power of the Stokes component is normalized to theory and shown as a function of the relative phase between modulation tones $\alpha$ at $(\omega - \omega_0)/(2\pi) = \Omega/(2\pi) = 20$ MHz.} 
\label{figSI1}
\end{figure}

According to our theory, isolator behavior is expected at the relative phases $\alpha/\pi \approx \pm 0.3$, where the reflection in both directions is strongly suppressed and where there is an asymmetry in transmission in opposite directions resembling the $S$-matrix of an ideal isolator
\begin{eqnarray}\label{SM_is}
&& S_{\rm isolator} = \begin{pmatrix}
0 & 0 \\
1 & 0 
\end{pmatrix}.
\end{eqnarray}
For the measurements at a modulation frequency $\Omega/(2\pi) = 20$\,MHz and modulation amplitude $A_m/(2\pi) = 30$\,MHz, we find an isolation of $10\log_{10}{(S_{21}/S_{12})} \sim 3.3$ dB and an insertion loss of $10\log_{10}{S_{21}} \sim 11$ dB at $\alpha/\pi \approx - 0.3$ as shown in Fig. \ref{figSI1}. 

While the insertion losses cannot be fully avoided both parameters could be improved with an increase of the modulation amplitude. 
%This aspect was not the main focus of this paper and could be explored further in the future. 
Moreover, adding more qubits to the device while keeping the effective distance between nearest neighbors at $\lambda/4$ and the relative phase $\alpha/\pi = 0.5$ between modulation tones would enable the creation of a topological isolator for both elastically and inelastically scattered light \cite{Lewenstein2014,Yuan:16,Poddubny2021Ratchet}.

\subsection*{Power dependence of directionality}
\begin{figure}[h]
\centering
\includegraphics[width=0.7\columnwidth]{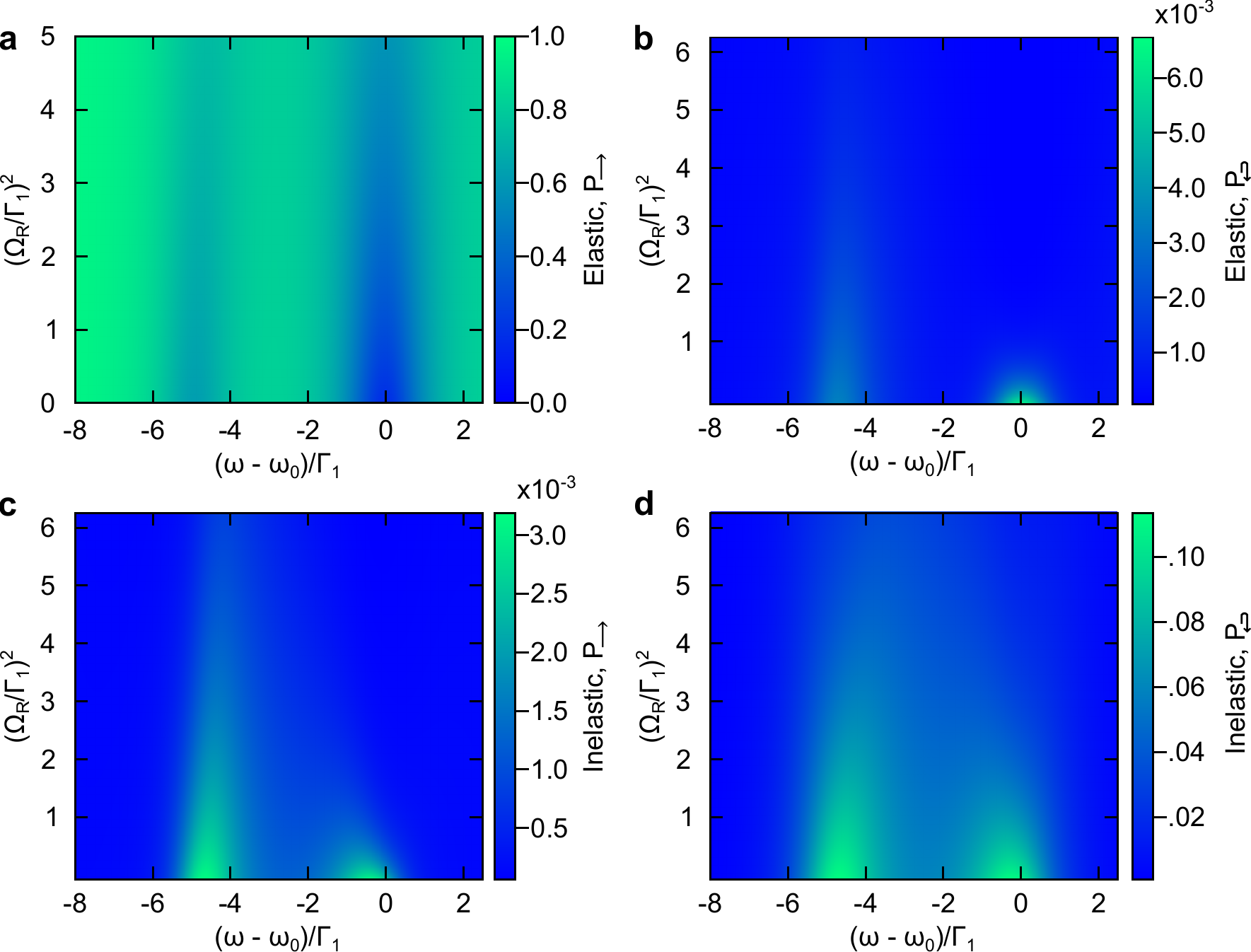}
\caption{\textbf{Dynamic range of directional scattering.} Calculated coherent elastic (\textbf{a, b}) and inelastic (\textbf{c, d}) scattering spectra as a function of normalized drive frequency $(\omega-\omega_0)/\Gamma_1$ and drive power $(\Omega_R/\Gamma_1)^2$ theoretically predicted for reflection (\textbf{a, c}) and transmission (\textbf{b, d}).
The calculation has been performed for the modulation phase difference $\alpha_2-\alpha_1=\pi$ and $\Omega=A_m=5\Gamma_1$.
} 
\label{figSI2}
\end{figure}

In this section we discuss the dependence of the directionality on the incident wave power. In order to calculate the scattering we use the master equation~\cite{Blais2013} for the density matrix $\rho$:
\begin{eqnarray}\label{eq:difsystem}
   \dot{\rho}=-i[H_1,\rho]+\sum_{j,k=1}^N\left(\frac{\Gamma_1}{2}\cos[\varphi(j-k)+\Gamma_2\delta_{j,k}\right)\left[2\sigma_j\rho\sigma_k^{\dagger}-\{ \sigma_k^{\dagger}\sigma_j ,\rho\}\right]
\end{eqnarray}
with the Hamiltonian
\begin{eqnarray}\label{eq:H2}
    H_1&=    \sum\limits_{j=1}^N[\omega_0+A_m(t)\cos(\Omega t+\alpha_j)]
    \sigma_j^\dag \sigma_j+\frac{\Gamma_1}{2} \sum\limits_{j,k=1}^N\sigma_j^\dag \sigma_k\sin(\varphi |j-k|)\\\nonumber
    &+ \sum\limits_{j=1}^N\frac{\Omega_R}{2} (\e^{-\rmi \varphi j-\rmi (\omega-\omega_0) t}\sigma_j^\dag +{\rm H.c.})\:.
\end{eqnarray}
The coherent reflection and transmission coefficients
that describe the backward (forward) scattering process with the frequency change $\omega \to \omega+n\Omega$ are then found as the Fourier transforms
\begin{gather}
    r_{n}=\frac{2}{\Omega_R}\lim_{T\to\infty}\frac{1}{T}\int\limits_0^T\rmd t\:
    \e^{-\rmi(\omega+n\Omega)t}\sum\limits_{j=1}^N[\mathop{\mathrm{Tr}}\nolimits \rho(t)\sigma_j^\dag] \e^{\rmi\varphi j}\:,\\
    t_{n}=\delta_{n,0}+\frac{2}{\Omega_R}\lim_{T\to\infty}\frac{1}{T}\int\limits_0^T\rmd t\:
    \e^{-\rmi(\omega+n\Omega)t}\sum\limits_{j=1}^N[\mathop{\mathrm{Tr}}\nolimits \rho(t)\sigma_j^\dag] \e^{-\rmi\varphi j}\:.
\end{gather}
The calculated forward and backward scattering spectra
$P_{\rightarrow}\equiv |t_n|^2 $ and $P_{\hookleftarrow}\equiv |r_n|^2$ are shown in Fig.~\ref{figSI2}. 
In the limit of  vanishing driving power this calculation yields the same results as the one presented in the main text. Namely,  the $\alpha=\pi$ phase difference between the modulation tones of first and second qubits corresponds to strong inelastic backscattering, see Fig.~\ref{figSI2}d. Increase of the Rabi frequency leads to a gradual suppression of the scattering directionality that persists up to $\Omega\approx 3\Gamma_1$. This can be interpreted as a result of the saturation of the qubit transition induced by a strong driving~\cite{astafiev2010resonance}.

\section*{Supplementary methods}

\subsection*{Modulation amplitude calibration}
\begin{figure}[h]
\centering
\includegraphics[width=.5\columnwidth]{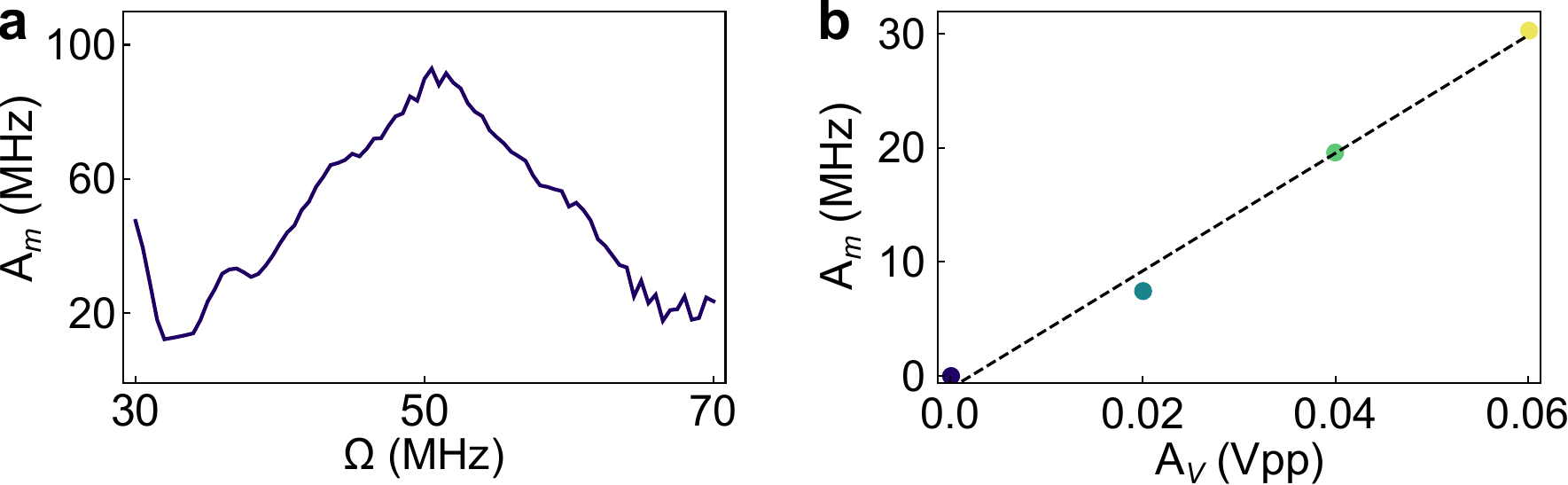}
\caption{\textbf{Modulation amplitude calibration. a, } Measured dependence of the modulation amplitude $A_m$ on the modulation frequency $\Omega$ at the fixed AWG amplitude $A_V = 0.2\,V_\text{pp}$. \textbf{b,} Measured dependence of the modulation amplitude $A_m$ on the applied AWG amplitude $A_V$ at the modulation frequency $\Omega = 20$ MHz.}
\label{figSI3}
\end{figure}

The measured frequency dependence of the modulation amplitude $A_m(\Omega)$ is shown in Fig.~\ref{figSI3}a. It is caused by the various low pass filter stages we use to suppress external flux noise on the bias input line. To calibrate the modulation amplitude $A_m$ we measure the transmission spectrum $|t_0|^2$ of a modulated qubit at the amplitudes $A_V$ from $0$ to $0.1$ $V_\text{pp}$ and for different modulation frequencies $\Omega$.

For each value of $\Omega$, we fit $A_m$ as a linear function of $A_V$ as shown in Fig.~\ref{figSI3}b. The linear fit is sufficient since we are far away from the flux sweet spot of the transmon qubit and $\partial^2 E_{01}/\partial \phi^2 \approx 0$. After repeating this procedure for each qubit separately where one is tuned to $\omega_0/(2\pi)=6.129$\,GHz and the other is far ($> 2$\,GHz) detuned, we can now calculate the required $A_V$ to result in the desired $A_m$ for all $\Omega$. 
%for different modulation frequencies.

\section*{Supplementary Figures}

%%%%%%%%%%%%%%%%%%%%%%%%%%%%%%%%%%%%%%%%%%%%%%%%%%%%%%%%%%%
%%%%%%%%%%%%%%%%%%%%%%%%%%%%%%%%%%%%%%%%%%%%%%%%%%%%%%%%%%%%

%%%%%%%%%%%%%%%%%%%%%%%%%%%%%%%%%%%%%%%%%%%%%%%%%%%%%%%%%%%
%%%%%%%%%%%%%%%%%%%%%%%%%%%%%%%%%%%%%%%%%%%%%%%%%%%%%%%%%%%%

%%%%%%%%%%%%%%%%%%%%%%%%%%%%%%%%%%%%%%%%%%%%%%%%%%%%%%%%%%%
%%%%%%%%%%%%%%%%%%%%%%%%%%%%%%%%%%%%%%%%%%%%%%%%%%%%%%%%%%%%

\end{document}